\documentclass[prodmode]{acmsmall} 

\usepackage[ruled]{algorithm2e}

\SetAlFnt{\small}
\SetAlCapFnt{\small}
\SetAlCapNameFnt{\small}
\SetAlCapHSkip{0pt}
\IncMargin{-\parindent}
\usepackage{subfigure}

\begin{document}

\markboth{M. Pasta et al.}{On Topology of Networks and Limitations of Community Detection Algorithms}

\title{On Varying Topology of Complex Networks and Performance Limitations of Community Detection Algorithms}
\author{M.Q. PASTA
\affil{KIET University\\ }
F. ZAIDI
\affil{City University of New York}
G. MELAN\c{C}ON
\affil{University of Bordeaux}
}

\begin{abstract}
One of the most widely studied problem in mining and analysis of complex networks is the detection of community structures. The problem has been extensively studied by researchers due to its high utility and numerous applications in various domains. Many algorithmic solutions have been proposed for the community detection problem but the quest to find the best algorithm is still on. More often than not, researchers focus on developing fast and accurate algorithms that can be generically applied to networks from a variety of domains without taking into consideration the structural and topological variations in these networks.

In this paper, we evaluate the performance of different clustering algorithms as a function of varying network topology. Along with the well known LFR model to generate benchmark networks with communities, we also propose a new model named Naive Scale Free Model to study the behavior of community detection algorithms with respect to different topological features. More specifically, we are interested in the size of networks, the size of community structures, the average connectivity of nodes and the ratio of inter-intra cluster edges. Results reveal several limitations of the current popular network clustering algorithms failing to correctly find communities. This suggests the need to revisit the design of current clustering algorithms that fail to incorporate varying topological features of different networks. 

\end{abstract}


%
%




%
%

\maketitle

\section{Introduction}\label{sec::introduction}

Many real world systems can be represented as networks depicting relationships and inter-dependencies in systems emerging from diverse domains such as social networks \cite{lin09a,namata16}, biological networks \cite{ibanez16}, geographical networks \cite{ducruet12} and economic network \cite{fagiolo13}.

Relationships in these networks naturally evolve with a non-uniform distribution of ties resulting in modular structures \cite{prat16,newman04c}. These modular structures are important as they not only help to understand the complex inter-connectivity among objects but are also vital towards better understanding, management and prediction of the entire system. These modular structures are usually called community structures, or simply clusters.

Even though the problem of finding community structures (or clusters) in networks has been widely studied by physicists, computer scientists, statisticians and mathematicians, there still remains a disagreement on a standard definition of communities in networks. The most widely accepted definition of a community is a high number of intra-cluster edges and low inter-cluster edges \cite{fortunato10,schaeffer07}. This translates to groups of nodes well connected to each other but sparsely connected to nodes from different clusters.

\begin{figure}[b!]
\centering
	\includegraphics[width=0.8\textwidth]{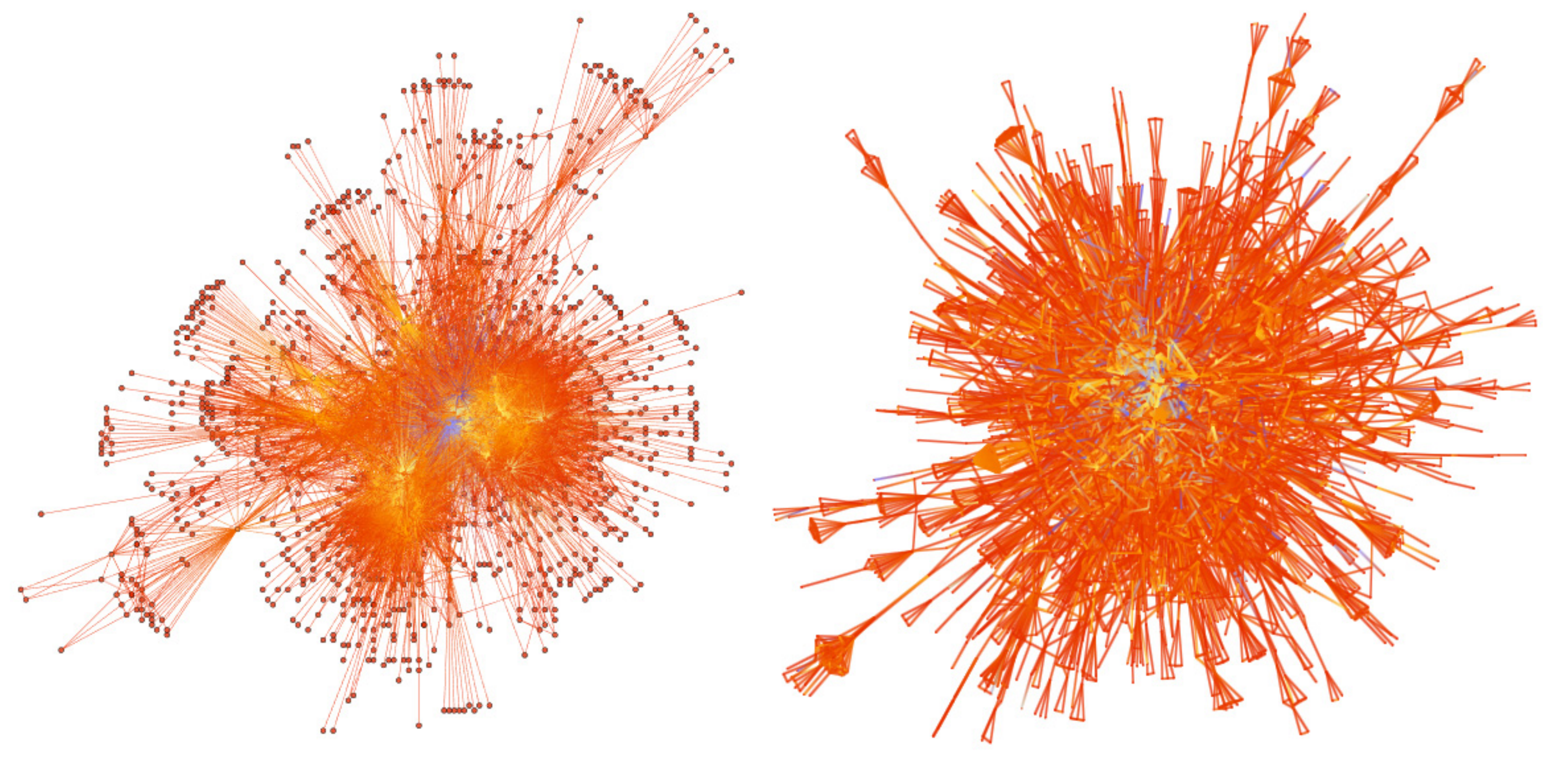}
	\caption{(Left) Air Transport Network (AT) where cities are represented by nodes and direct air flights link cities. (Right) Co-Authorship network (CA) where authors are represented by nodes and co-authoring an article links authors.}
	\label{fig::examples}
\end{figure}

Real world networks usually exhibit small world \cite{watts98} and scale free \cite{barabasi99} characteristics implying some structural similarity among them, but there are inherent topological and structural features that make these networks quite different from each other, specially for the clustering problem. One obvious reason is that these networks emerge from a variety of different domains introducing structural differences in networks such as, the average connectivity of nodes which in turn affects the performance of clustering algorithms. Furthermore, algorithms usually try to optimize specific properties such a modulariy \cite{clauset04} or distances \cite{pons04} disregarding the different structural variations present in these networks. For example, networks of air traffic \cite{rozenblat08} usually contain a few thousand nodes representing cities with airports as compared to networks obtained from online social networking website \cite{cheng12d}\cite{mislove07} which at times contain millions and billions of users indicating the huge difference in network sizes. Similarly, users are on average connected to hundreds of friends through online social networking websites as compared to protein interaction networks \cite{wagner03} which usually have an average degree, even below 2 or 3. Figure \ref{fig::examples} shows two real networks, the Air Transport Network (AT)\cite{rozenblat08} and the Geometry Co-Authorship Network (CA)\cite{geometrynetwork06}. Some basic network metrics for these two networks are shown in Table \ref{tbl::one} where their topological differences can be easily identified. These examples motivate the need to identify these structural differences as an important characteristic to be accounted for while designing and applying clustering algorithms on networks. More recent developments in online social networks have generated networks with billions of nodes and edges, which present new challenges both in terms of scalability and more importantly, accuracy to determine communities in these huge sized networks. 

\begin{table}%
\centering
\tbl{Metrics for Two Real World Networks Highlighting Topological Variations\label{tbl::one}}{%
\begin{tabular}{|c|l|l|}
\hline
\textbf{Metrics} 		          & \textbf{Air Transport (AT)} 			& \textbf{Co-Author (CA) }   \\\hline

Number of Nodes  (N)                  & 1540    & 3621	\\\hline
Number of Edges  (E)                 & 16523   & 9461  \\\hline         
Node-Edge Ratio  ($\mathit{R}$)                 & 10.7    & 2.6   \\\hline
Clustering Coefficient (CC)            & 0.26    & 0.22  \\\hline
Average Path Length (APL)              & 2.93    & 5.31  \\\hline
Power-Law Coefficient ($\gamma$)            & 2.68    & 2.45  \\\hline
Maximum Node Degree (Max-D)            & 487     & 102   \\\hline
Number of Nodes with Degree 1 (D-1)    & 332     & 741   \\\hline
Highest K-core (Max-K)                   & 44      & 21    \\\hline

\end{tabular}}
\begin{tabnote}%
\Note{}{The metrics show the variations in the two real world networks. $\mathit{R}$ is clear indication of how different the average node connectivity is in the two networks. APL shows on average, the nodes of AT are much closer to each other than the nodes of CA network. Max-D of AT shows that the high degree nodes are very well connected to the entire network as compared to the CA network. Max-K and D-1 of AT and CA networks show the variation in the distribution of the connectivity among low and high degree nodes in the two networks.} 
\end{tabnote}%
\end{table}%

In this paper, we focus on structural and topological characteristics of networks and analyze how they affect the performance of clustering algorithms. More specifically, we are interested in studying the effects of the size of networks in terms of number of nodes, number and size of communities present in a network, the average connectivity (or average degree) of nodes in a network and the ratio of inter-intra cluster edges, also referred to as the mixing of clusters, in a network. The goal of this study is not to find the best algorithm, but to expose the weaknesses of clustering algorithms in general, when tested against various topological features of networks.

The paper is organized as follows: Section \ref{sec::related} presents the related work and Section \ref{sec::experimental} describes the experimental setup. In Section \ref{sec::results}, we present the main findings and discuss their possible implications in the study of community detection algorithms. Finally, we conclude in Section \ref{sec::conclusions} where we provide possible future research directions.

\section{Related Work}\label{sec::related}

The related work is grouped into three sub-sections each addressing an important component of our experimental setup. First we discuss benchmark graphs which provide the basis for our experiment, second we discuss several network generation models used to generate networks with community structures and the third sub-section provides a brief account of works where clustering algorithms are compared.

\subsection{Benchmark Graphs}

Although the problem of community detection has been widely studied, the associated problem of generating standard benchmark datasets to evaluate the quality of clustering algorithms has not attracted much interest. One of the earlier works in this direction was by Girvan and Newman \cite{girvan02} usually referred to as the GN benchmark. They used a network of 128 nodes divided into four communities of 32 nodes each with each node having approximately the same node degree. A parameter was used to control the intra-cluster and inter-cluster edges. A number of drawbacks were identified in this benchmark \cite{lancichinetti08}, most notably that all nodes of the network have the same degree and all the communities have the same number of nodes in them. 

Most networks found in the real world have a non-uniform degree distribution, often following power-law \cite{barabasi99}. Furthermore, the community sizes in these real networks also follow a power-law \cite{palla05}\cite{guimera03}\cite{xie07} which in turn justifies the drawbacks identified in the benchmark proposed by \cite{girvan02}. Thus Lancichinetti et al. \cite{lancichinetti08} proposed a new benchmark called the LFR benchmark which has a number of interesting features. The degree distribution of the generated networks follow power-law and the average degree can be adjusted as required. The distribution of the community sizes also follow power-law and can be parametrized between a minimum and maximum value. Each node can have a fraction of intra-cluster and inter-cluster edges which are also controlled by a mixing parameter. 

Lancichinetti and Fortunato \cite{lancichinetti09} also proposed methods to generate benchmarks for directed and weighted networks as well as networks with overlapping community structures but in this study, we limit our analysis to undirected, unweighed graphs with hard and flat community structures.

\subsection{Network Models with Community Structures}

There exists a number of algorithmic models to generate networks in order to mimic real world graphs. The primary objective of these networks is to reproduce networks similar to real world networks with structural properties such as low average path lengths, high clustering coefficients, degree distributions following power-law and the presence of community structures. Although not used as algorithms to generate benchmark graphs to evaluate clustering algorithms, they provide a number of options for such a study. We review some of these algorithms below.

Li and Chen \cite{li06a} propose a model to generate weighted evolving networks incorporating three types of power-law distributions, first on the node degree, second on link weights and third on node strengths along with the presence of clear communities. Xie \textit{et al.} \cite{xie07} introduced an evolving network model to generate clustered networks with the cumulative distribution of community sizes following power-law. The basic idea is when new connections between communities are added, or a new node to an existing community is added, communities with larger sizes are selected preferentially. Zhou \textit{et al.} \cite{zhou08} worked with two important topological characteristics, first, the dense intra-cluster connections as compared to inter-cluster connections and second, size of communities following a power-law just as \cite{xie07} proposed. Based on these characteristics, they proposed a weighted growing model with power-law distributions of
community sizes, node strengths, and link weights. 

Kumpula \textit{et al.} \cite{kumpula09} used the concepts of cyclic closure and focal closure from sociology to propose a model to generate a weighted network with communities. New links are created preferably through strong ties which make these links stronger. Xu \textit{et al.} \cite{xu09} presented a model with community structures using the idea of local events using three processes, adding new intra-community nodes, new intra-community links or new inter-community links. The model uses preferential attachment mechanism resulting in power law degree distribution.

Moriano and Finke \cite{moriano13} proposed a model to generate networks with groups of nodes densely connected to each other and sparsely connected with other nodes. The model attempted to explain networks with extended power law degree distributions and clustering coefficients that do not diminish for huge size networks. The connectivity of new nodes probabilistically choose nodes of same type to form community structures.

Zaidi \cite{zaidi13a} proposed a model to generate clustered networks with high clustering coefficient and low average path lengths. The author demonstrate that clustered networks can be generated from completely random graphs by introducing some order, which is a contrasting approach to the famous model of Watts and Strogatz \cite{watts98}. The model is further extended to generate clustered networks where communities are randomly connected to each other. Sallaberry \textit{et al.} \cite{zaidi13b} also proposed a static network generation model with community structures i.e\ the number of nodes added at the start remain the same throughout the algorithm and only edges are rewired to create communities. The model is probabilistic and increases the edge connectivity among nodes closer to each other and reduces edges among nodes far apart in the network. 

Pasta \textit{et al.} \cite{pasta13a} recently proposed a tunable network generation model with community structures whose flexibility allows it to generate a variety of networks with varying structural properties. The authors focus on three structural features of community structures generated through this model, the degree distribution within each community follows power-law, high clustering coefficient of nodes within each community, and each community can be further divided into sub-communities.

\subsection{Comparative Analysis of Clustering Algorithms}

Danon \textit{et al.} \cite{danon05} studied network clustering algorithms and compared the performance of 16 different algorithms through empirical analysis using benchmark datasets generated by \cite{girvan02}. The authors also compared the asymptotic complexity of different clustering algorithms to conclude that small datasets may be clustered using slow but more accurate methods such as \cite{guimera04} and for large networks, the quality might be compromised with a faster algorithm requiring less computational effort.

Lancichinetti and Fortunato \cite{lancichinetti09a} comapared the performance of 12 different clustering algorithms using the GN and the LFR benchmarks. The authors focus on mixing parameter to perform a comparative analysis for different values of network sizes (1000 and 5000) and different community sizes (between 10 to 50 and 20 to 100) in an attempt to identify the algorithms that perform well on these benchmarks. The authors also performed a comparative analysis for directed-unweighted graphs, undirected-weighted graphs and undirected-unweighted graphs with overlapping community structures. As compared to this study, we also use the LFR benchmark but vary parameters with a different perspective specially varying the average degree of the graphs and large variation in the size of clusters generated. 

Orman and Labatut \cite{orman09} compared five community detection algorithms to study the behavior of mixing parameter and evaluate the performance of community detection algorithms concluding that walktrap and spinglass methods perform better. They also conclude that the average and maximum degree of nodes have a strong joint effect on the results of community detection algorithms. They only consider relatively small size networks (nodes = 1000) varying maximum node degree and the average node degree.

Leskovec et al. \cite{leskovec10} studied various network and common objective functions optimized to detect communities with the aim to understand the structural properties of identified clusters by different methods with the overall motive to find the best suited algorithm for specific applications. They conclude that the performance of community detection algorithms varies for certain classes of networks for example, communities of larger size tend be less dense.

Orman et al. \cite{orman11} compared different community detection algorithms based on community-centered properties such as  heterogeneity of community sizes (few large and many small communities), embeddedness (ratio of neighbors in the same community to total neighbors) and density (ratio of links within a community to total links possible among the nodes of a community). The performance of different community detection algorithms was quantitatively compared to these community-centered properties. This differs from past approaches where objective measures were used as quality measures (such as Modularity \cite{newman04}). Important findings include the algorithms behave diversely when the number of communities vary in the network and the inability of Normalized Mutual Information (NMI) to correctly quantify community detection algorithms.

Sousa and Zhao \cite{sousa14} empirically compared different clustering algorithms using synthetic and real networks. The objective of their work was to evaluate the performance of different algorithms. They found that the performance of some algorithms is effected by the number of nodes in a network. The artificial networks they generated were with a maximum node size of 100 which makes it difficult to generalize the results to large networks with hundreds and thousands of nodes. They use Modularity Q \cite{newman04} in part, to evaluate the quality of clustering algorithms which is shown to have problems with underestimation \cite{fortunato07} and overestimation \cite{kehagias13} of the number of communities when maximizing modularity.

A more recent work in evaluating network clustering algorithms was performed by Wang \textit{et al.} \cite{wang15}. They proposed a procedure-oriented framework for benchmarking to evaluate different community detection approaches in an attempt to find better algorithms. Although their objectives were clearly different from what we have proposed in this paper, one of their findings also pointed towards the varying performance of algorithms for different networks.

\section{Clustering Algorithms}\label{sec::clustering}

In this section, we discuss several clustering algorithms commonly used to detect community structures in networks. An exhaustive  review of all the algorithms present in the literature is out of our scope, but we have tried to select algorithms that are popular in term of their wide application and where their implementation is available in graph clustering and analysis software.

One of the earliest methods of the new generation of algorithms for network clustering is that of Girvan and Newman \cite{girvan02}. The algorithm iteratively calculates the betweenness centrality of edges, removes the edge with the highest value and repeats the process to find disconnected set of nodes. Since the algorithm performs a global calculation of betweenness centrality in every step, it is very slow and not applicable on large networks. The algorithm stops for optimal values of a quality metric called Modularity (Q) which measures the fraction of the intra-community edges minus the expected value of edges randomly distributed among vertices of the same quantity in a network. Radicchi et al. \cite{radicchi04} proposed an algorithm based on similar idea. Instead of calculating edge betweenness to determine the edges that lie in between communities, they calculate edge clustering coefficient, which is a local metric to calculate the ratio of the actual triads that are present around an edge to the maximum number of triads possible. Auber et al. \cite{auber03} also proposed an algorithm where they used cycles of length 3 and 4 to identify edges between different communities.

A number of local clustering algorithms have been proposed that try to optimize modularity locally such as \cite{clauset04}\cite{guimera05a}\cite{wakita07}\cite{blondel08}. These methods in essence try to agglomerate nodes which result in high modularity. For example, the method of Clauset et al. \cite{clauset04} (Fastgreedy Clustering), starts from a set of isolated nodes and iteratively add edges such that the increase in modularity is the highest. The method proposed by \cite{guimera05a} performs an exhaustive optimization of modularity as compared the method proposed by \cite{clauset04} and is thus expected to perform better in terms of modularity achieved but is slower than the implementation of \cite{clauset04}. 

Another algorithm which has gained a lot of popularity is the algorithm proposed by Blondel et al. \cite{blondel08} (Multilevel Clustering). This is a multistep approach which first tries of optimize modularity locally by merging nodes, and then groups these nodes together to form supernodes, which in turn generates a new graph. The process is repeated until all the nodes are grouped together. Modularity is then used to find the right number of clusters from the obtained dendrogram.

Apart from modularity optimization, a number of dynamic processes have also been used to detect community structures in networks. Most notably, random walks belong to this classification with algorithms such as \cite{pons04}\cite{zhou04}\cite{rosvall08}. Pons et Latapy \cite{pons04} (Walktrap) propose a method which uses random walks to calculate distances between different nodes, based on which nodes are grouped together to form clusters. Nodes are grouped together using Ward's Method and Modularity is used to select the best partition of the resulting dendrogram. Rosvall and Bergstrom \cite{rosvall08} use random walks and try to compress the information of this dynamic process minimizing description length of the random walk to obtain clusters. Dongen proposed the Markov clustering (MCL) \cite{dongen00} algorith based on simulation of stochastic flow using Markov matrices. The algorithm iteratively applies two processes, expansion and inflation, resulting in a matrix representing a disconnected graph. The connected components are then grouped as clusters. 

Rosvall and Bergstrom \cite{rosvall08} (Infomap clustering) model the problem of community detection as problem of compressing information of a random walk taking place on the network. The algorithm optimizes the minimum description length of a random walk. Raghavan \textit{et al.} (Label propagation clustering) \cite{raghavan07} also proposed an algorithm based on an iterative dynamic process. Each node is initialized with a unique label and iteratively, nodes adopt the label which is the most frequent label of their neighbors. 

A completely different approach from these network based methods is the use of network layout methods to determine community structures. One popular mapping quality function VOS \cite{eck10} is used to propose VOS community detection algorithm \cite{waltman10} which simply optimizes the VOS mapping instead of the modularity function. The advantage is that it provides a good alternate to the Q modularity used in different community detection algorithms.

Another class of algorithms is based on the spin models, popular in statistical mechanics with a number of clustering algorithms using this analogy \cite{reichardt04,reichardt06}. The idea is based on a Hamiltonian which is derived from the principal that nodes that are linked together should belong to the same community and nodes which are not linked to each other should belong to different communities \cite{reichardt06a}. If Potts spin variables are assigned to the vertices of a network, and the interactions are between neighboring spins, structural clusters can be found from spin alignment of the system. Spins of nodes within clusters are similar and different across clusters where the idea is to maximize Potts energy.

More detailed information can be found in the respective citations of these algorithms. More literature and clustering algorithms can be found in these surveys \cite{newman04c}\cite{schaeffer07}\cite{fortunato10}. 

\section{Experimental Setup}\label{sec::experimental}

In this section, we describe the models used to generate benchmark graphs and the community detection algorithms used for experimentation.

\subsection{Becnhmark Graphs}\label{sec::bench}

We used two network models to generate benchmark graphs with community structures. The well known LFR model \cite{lancichinetti08}(see \cite{websitelfr} for implementation) which is the most widely used model (see \ref{sec::implementation}  for references for the implementation of LFR model), and a self designed model which we call naive scale-free clustering (NSC). 

The NSC model is inspired from the author's previous work \cite{zaidi13a} and uses a naive approach to build networks with community structures. For simplicity, consider the example of generating a graph with only two communities. As a first step, we used the BA model \cite{barabasi99} to generate two separate scale free networks of desired sizes (number of nodes, average degree). These two networks represent two communities and the nodes are labelled accordingly. Second, we connect the two networks by randomly selecting one node each from the two communities and adding an edge connecting the two networks. This step is repeated until the desired mixing is introduced. As a result, we obtain a network of two communities loosely connected to each other. The two communities have different network size and their degree distribution follows power-law. The algorithm is presented as Algorithm \ref{alg:nsc}. Existing models (discussed in section \ref{sec::related}) were not used due to their inherent complexity in controlling the topology of the generated networks. 

\begin{algorithm}[t]
\SetAlgoNoLine

\KwIn{$C[1...{C_L}]$ where C is an array with size of communities, ${L}$ is the Number of communities, $<k>$ is the average degree of nodes and $\mu$ is the mixing parameter.}


\KwOut{Graph $G$ with community structures.}

\For{$i = 1$ to $L$}
{
	$G[i]$ = $generateBarabasiGraph(size = C[i], avg. deg. = <k>)$ \; 
	$E_{\mu}[i] = |Edges(G[i])| * \mu $ \; 
}

 \While {$ (\sum_{i=1}^{L} E_{\mu}[i]) > 1   $} 
 {
      	$c_1 = getCommunity(C)  \{| E_{\mu}[c_1]>1\}  $            \; 
      	$c_2 = getCommunity(C)  \{| c_1 \neq c_2 \cap E_{\mu}[c_2]>1\} $  \; 
						
      	$n_1 = getRandomNode(c_1)$\;       	
      	$n_2 = getRandomNode(c_2)$\;
      	
      	$createEdge(n_1,n_2)$\;
      	
      	$E_{\mu}[c_1] = E_{\mu}[c_1] - 1$\;
		$E_{\mu}[c_2] = E_{\mu}[c_2] - 1$\;
 	
}
\caption{Naive Scale-free Clustering}
\label{alg:nsc}
\end{algorithm}

The use of two benchmark models to determine clustering quality ensures that the results can be generalized independently of the underlying network generation model. Both models allowed to tune input parameters and generate networks with desired structural and topological features. The four controlling parameters used to study the effects of network topology for clustering algorithms are described below:

\textbf{Network Size ($n$):} determines the size of the network in terms of its number of nodes. Networks were generated with three different sizes in terms of number of nodes $10^3$, $10^4$ and $10^5$.

\textbf{Average Node Degree ($k$):} controls connectivity for each node on average in the network. We generated networks with average degree $3$, $5$ and $10$.

\textbf{Mixing Parameter ($\mu$):} is on average, the fraction of edges a node has with its neighbors in the same community as opposed to its neighbors which are not in the same community. We used the values of $0.2$, $0.5$ and $0.8$.

\textbf{Minimum-Maximum Number of Clusters:} This pair of parameters determined the number of clusters ($\varsigma$) in a network. Three different ranges were used: Many clusters (many clusters in the network with small community sizes), Middling clusters (moderate number of clusters with moderate sized communities) and Few clusters (a few clusters with large sized communities). These three ranges were used for each of the three network sizes (networks with 1000, 10000 and 100000 nodes). For networks with 1000 nodes, the ranges for the three sizes in number of nodes are $20-50$, $100-150$ and $200-300$ giving us many small clusters for the smallest range ($20-50$) and few large clusters for the largest range ($200-300$). Similarly, for networks of $10,000$ nodes, the ranges used are $200-500$, $1000-1500$ and $2000-3000$, and for networks of $100,000$ nodes, the ranges are $2000-5000$, $10000-15000$ and $20000-30000$. 

The various combinations of these four parameters produced $81$ unique networks each for the two benchmark models, and the results were averaged over five instances of each of these networks.

\subsection{Community Detection Algorithms}\label{sec::community}

We used seven different clustering algorithms. The algorithms used cover a wide spectrum of clustering methods and techniques commonly used in graph clusterings. This is to avoid any bias created by a specific class of algorithms or optimization strategy. The selected algorithms are Fast Greedy, Multilevel, MCL, InfoMap, Label Propagation, VOS and Spinglass Clustering. (See \cite{clusteringsoftware} for implementation details of clustering algorithms.)

\subsection{Quantitative comparison of Clustering Algorithms:}\label{sec::quantitative}

A number of cluster similarity measures have been proposed in the recent past to quantitatively compare the community structure of two clusters, although some of these measures have drawbacks and are at times ill defined \cite{lancichinetti09}. We used the most widely accepted method for comparison, called the Normalized Mutual Information (NMI) first proposed by Danon \textit{et al.} \cite{danon05}. Given two partitional structures of a network, calculation of NMI returns a value in the range between 0 and 1 where 1 suggests perfect similarity and values close to 0 indicate high dissimilarity in the two partitions.

\begin{figure}
\subfigure[$\mu = 0.2$]{\centering \includegraphics[width=0.16\textwidth]{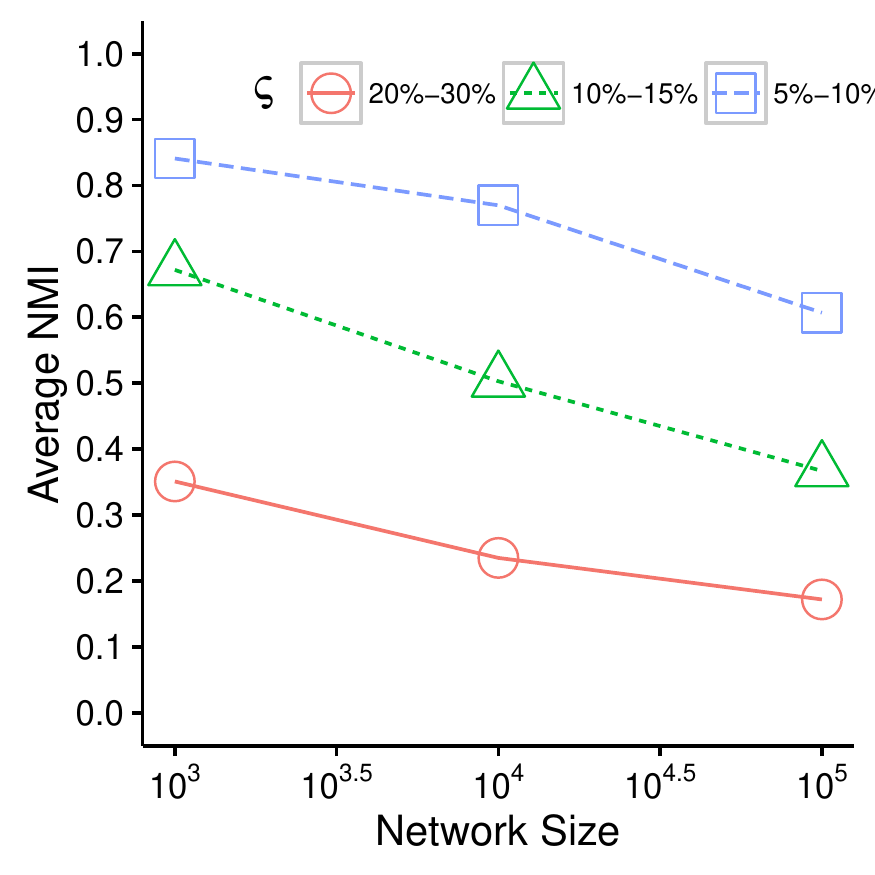}}
\hfill
\subfigure[$\mu = 0.5$]{\centering \includegraphics[width=0.16\textwidth]{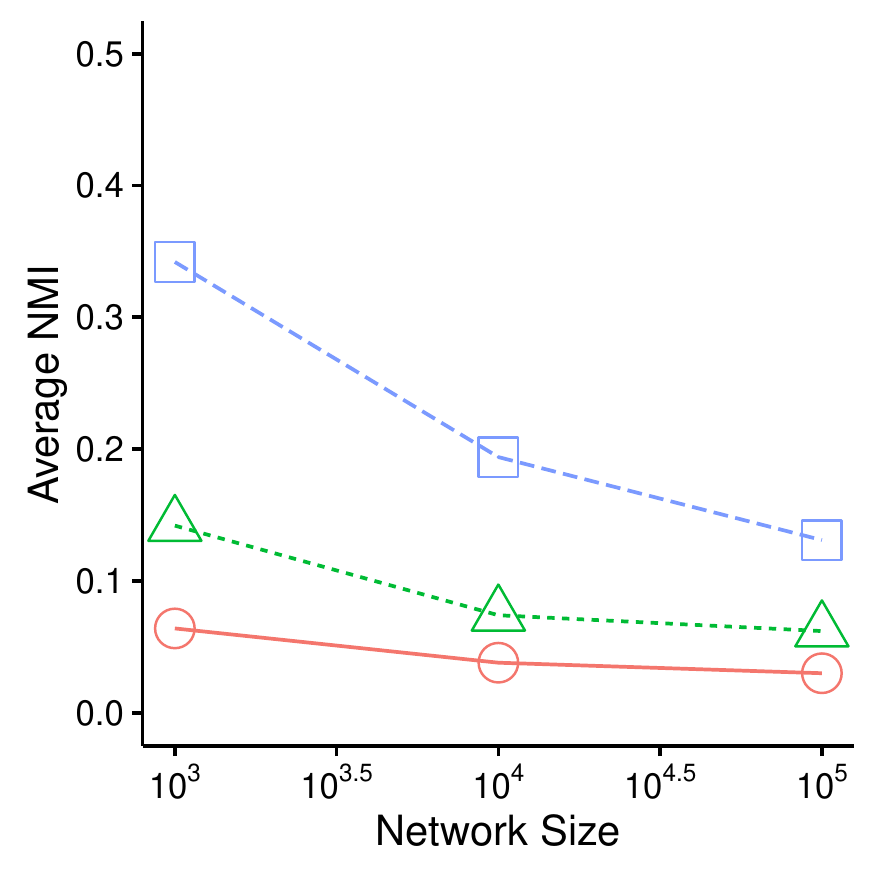}}
\hfill
\subfigure[$\mu = 0.8$]{\centering \includegraphics[width=0.16\textwidth]{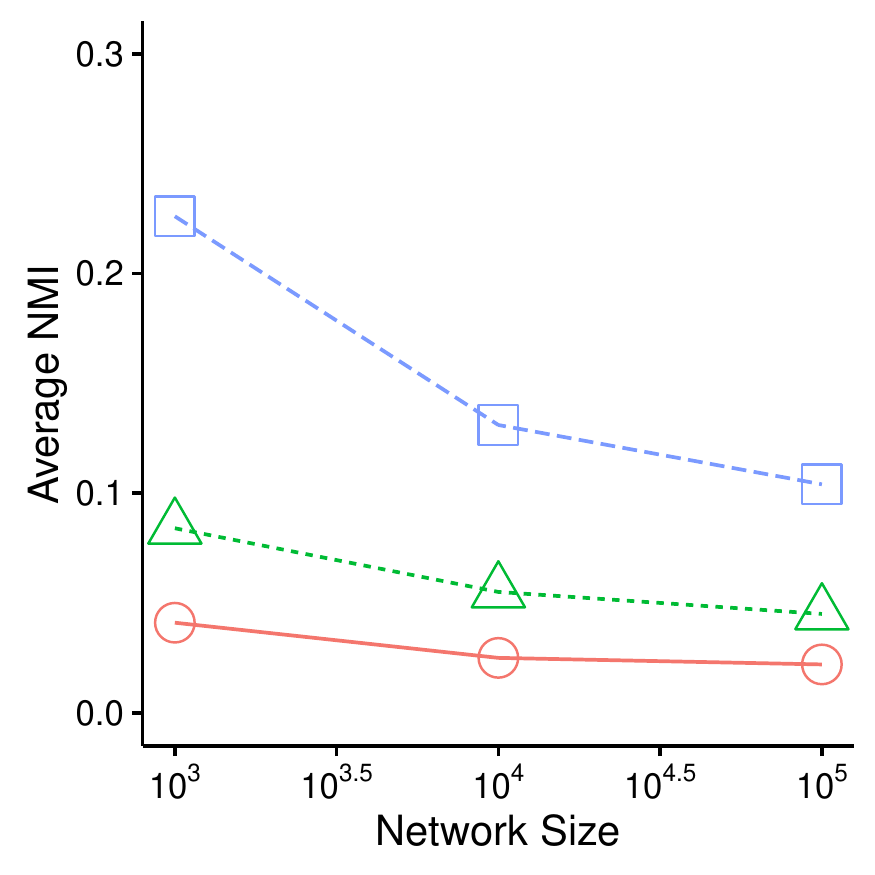}}
\subfigure[$\mu = 0.2$]{\centering \includegraphics[width=0.16\textwidth]{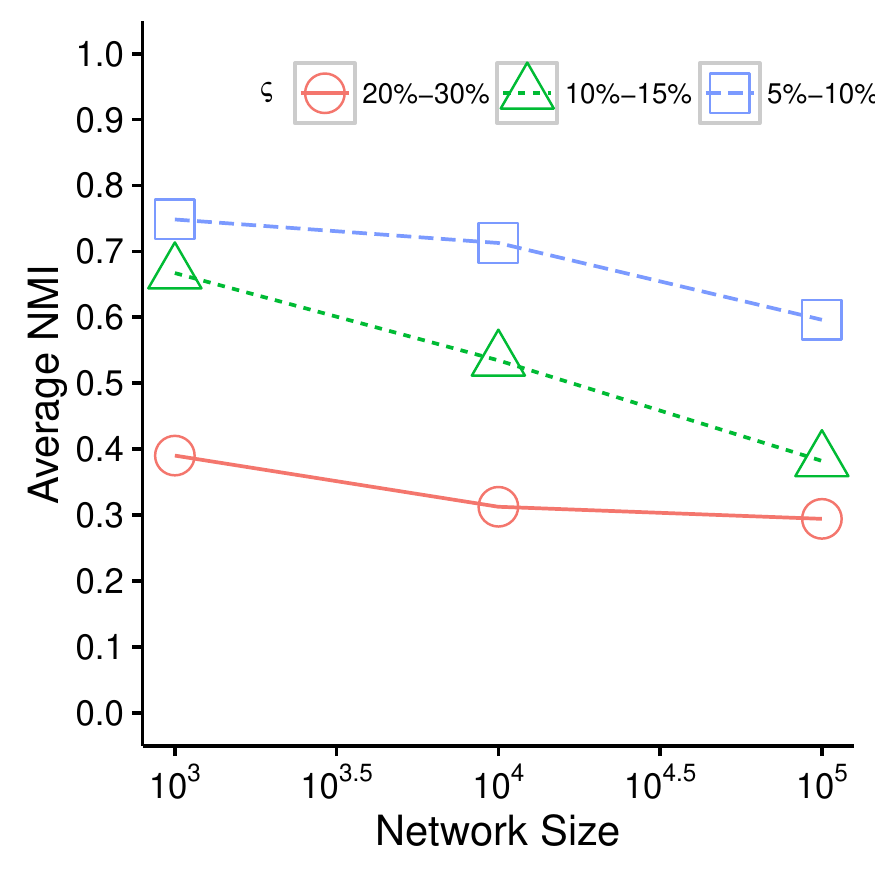}}
\hfill
\subfigure[$\mu = 0.5$]{\centering \includegraphics[width=0.16\textwidth]{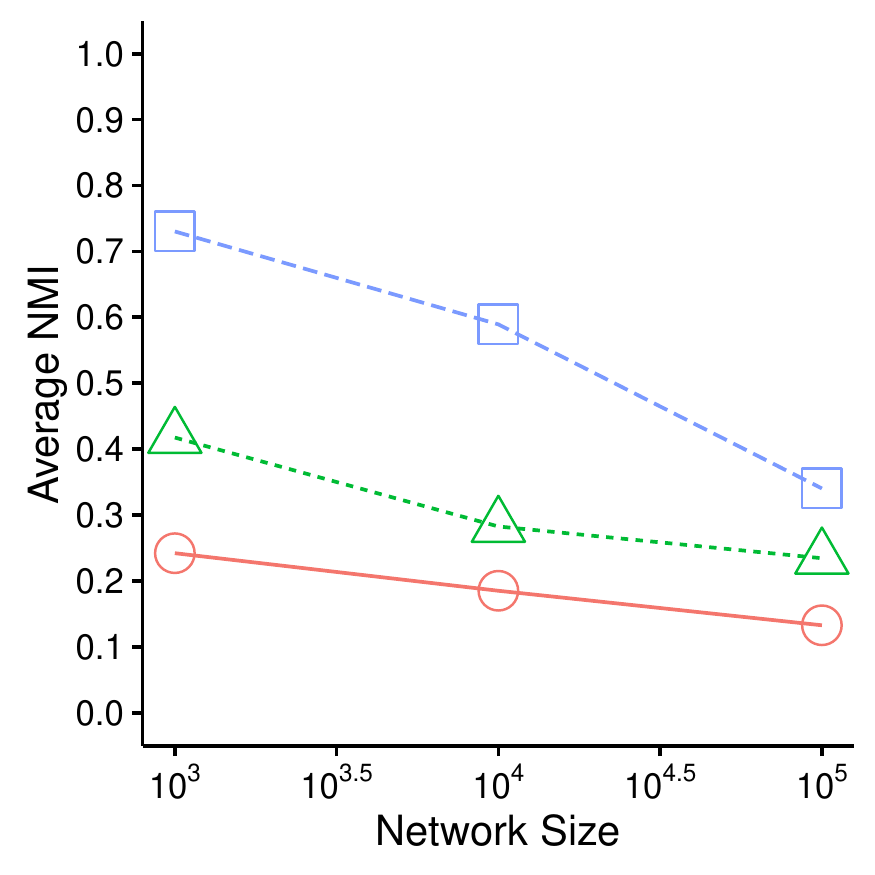}}
\hfill
\subfigure[$\mu = 0.8$]{\centering \includegraphics[width=0.16\textwidth]{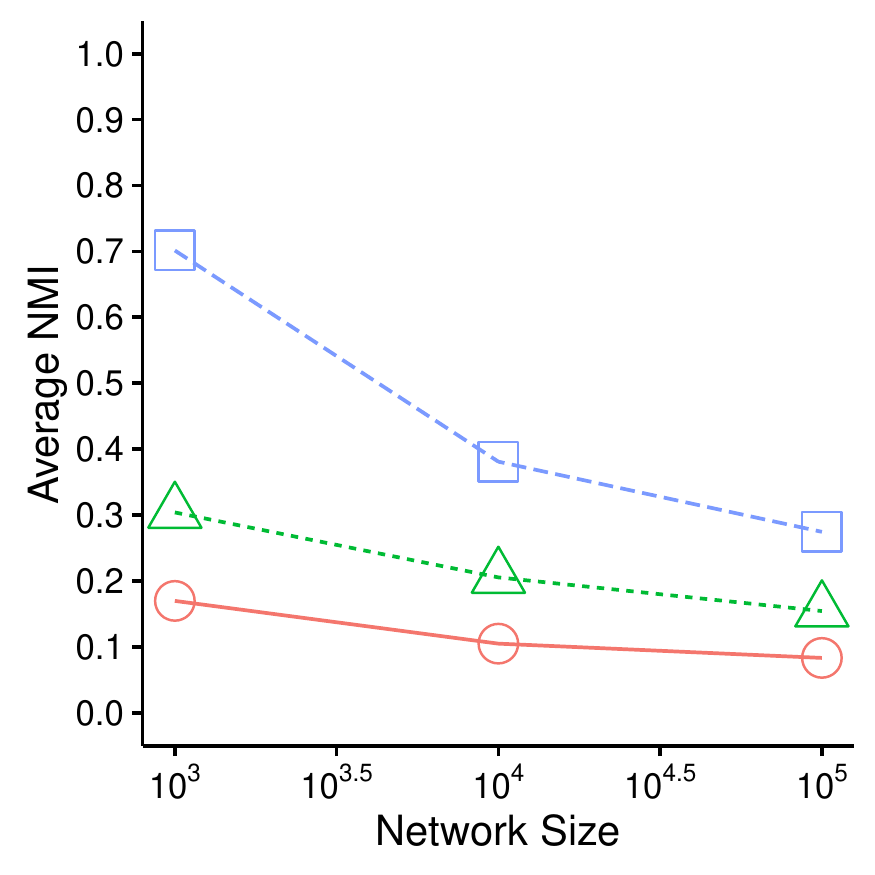}}
\caption{Results for LFR (a,b,c) and NSC (d,e,f) models. The average normalized mutual information as a function of network size($n$) with $n=(10^3,10^4,10^5)$. (a,d) Mixing Parameter $\mu=0.2$, (b,e) $\mu=0.5$ and (c,f) $\mu=0.8$ and different cluster sizes $ \varsigma =(5-10\%, 10-15\%, 20-30\%)$. The Clustering Algorithms perform poorly in two cases, when the size of the networks increases (along x-axis) and when there are a few large sized clusters in a network (represented by round markers).}%
\label{fig::size}
\end{figure}

\begin{figure}
\subfigure[Network of $10^{3}$ nodes]{\centering \includegraphics[width=0.24\textwidth]{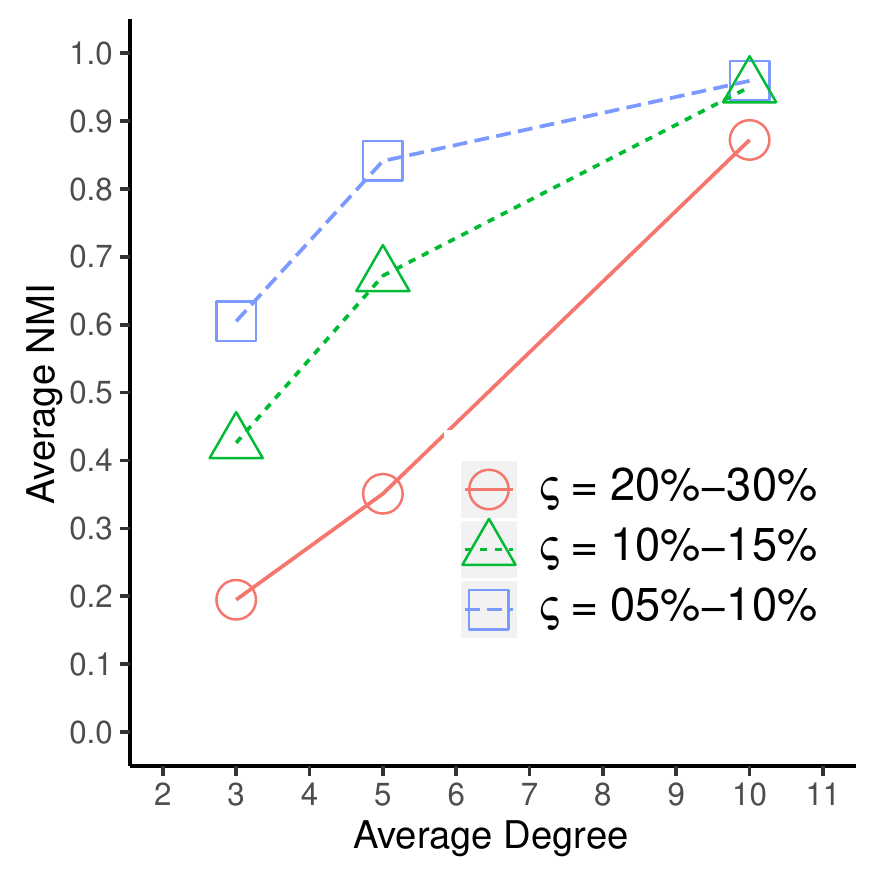}}
\hfill
\subfigure[Network of $10^{5}$ nodes]{\centering \includegraphics[width=0.24\textwidth]{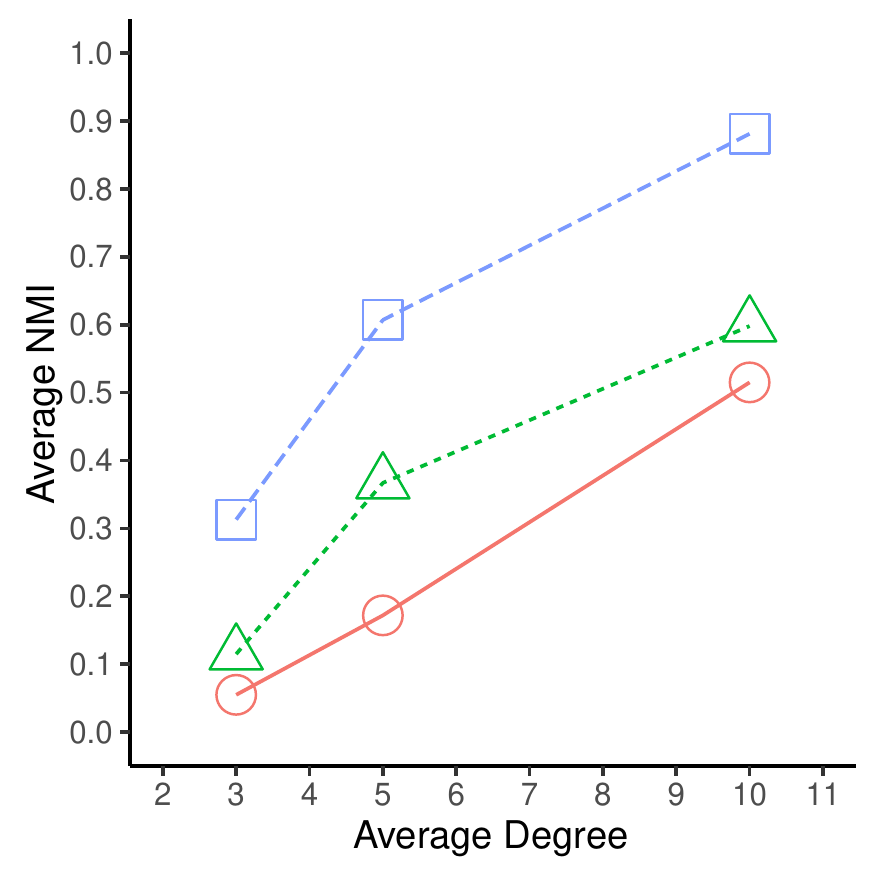}}
\subfigure[Network of $10^{3}$ nodes]{\centering \includegraphics[width=0.24\textwidth]{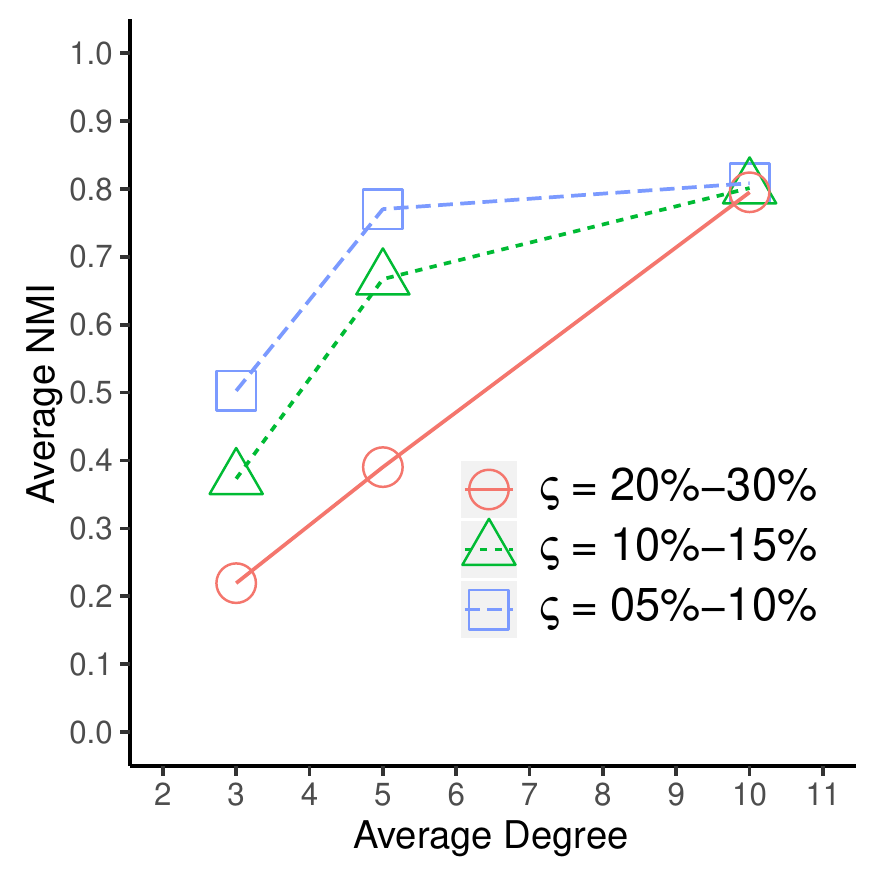}}
\hfill
\subfigure[Network of $10^{5}$ nodes]{\centering \includegraphics[width=0.24\textwidth]{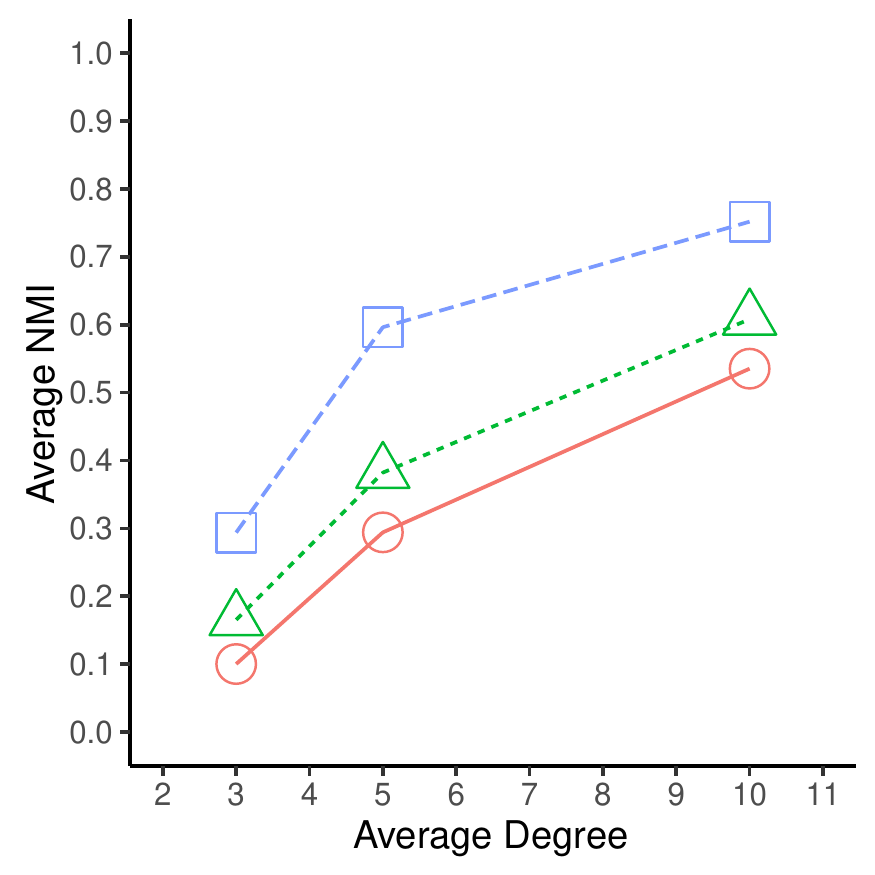}}
\caption{Results for LFR (a,b) and NSC (c,d) models. The average normalized mutual information as a function of average degree($\langle k \rangle=(3,5,10) $). (a,c) $10^{3}$ nodes (b,d) $10^{5}$ nodes and different cluster sizes $ \varsigma = (5-10\%), (10-15\%), (20-30\%) $ represented by different line markers.  The algorithms perform poorly for networks with low average degree regardless of network sizes and cluster sizes.}
\label{fig::cluster}
\end{figure}

\begin{figure}
\subfigure[$\varsigma = 2-5\%$]{\centering \includegraphics[width=0.16\textwidth]{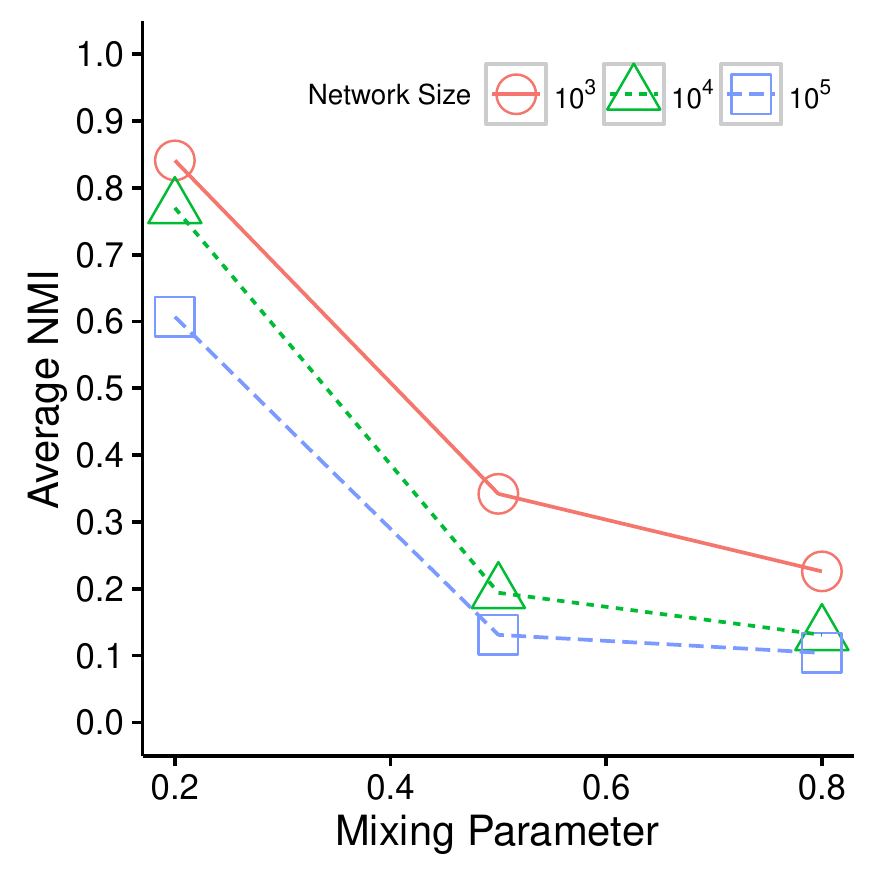}}
\hfill
\subfigure[$\varsigma = 10-15\%$]{\centering \includegraphics[width=0.16\textwidth]{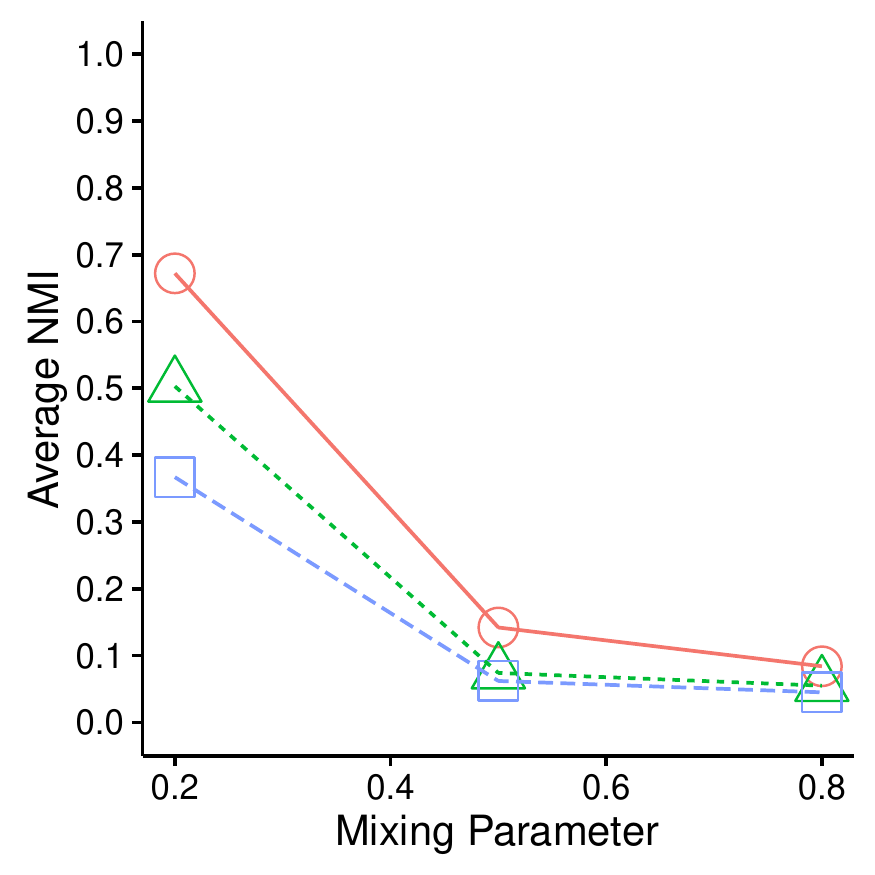}}
\hfill
\subfigure[$\varsigma = 20-30\%$]{\centering \includegraphics[width=0.16\textwidth]{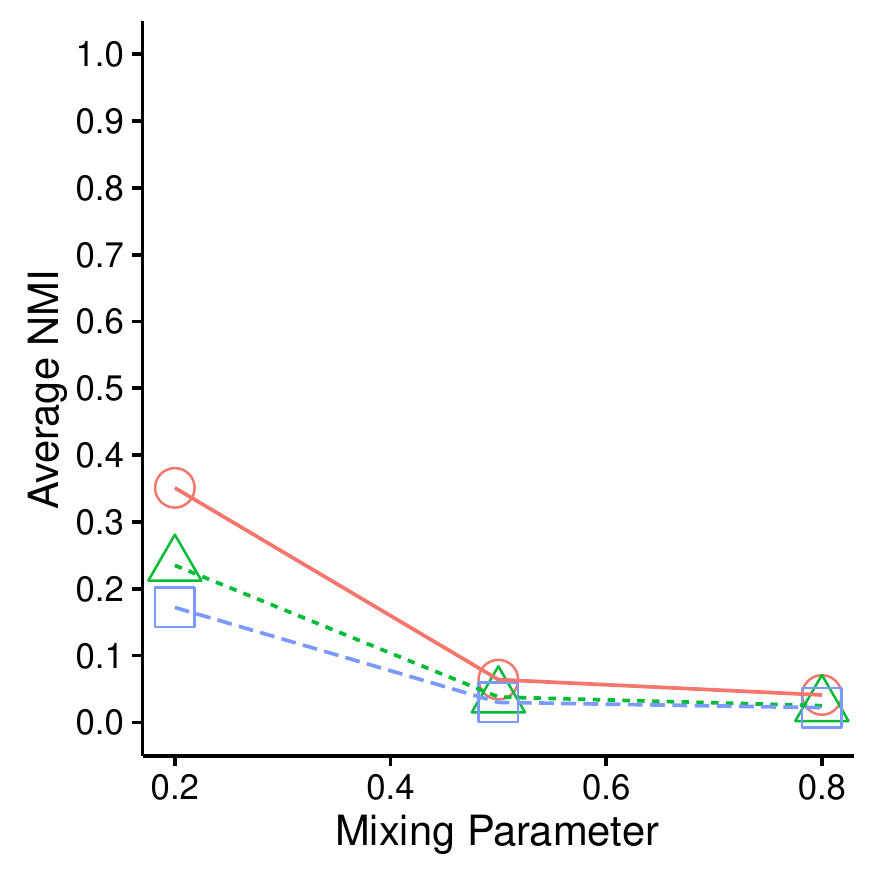}}
\subfigure[$\varsigma = 2-5\%$]{\centering \includegraphics[width=0.16\textwidth]{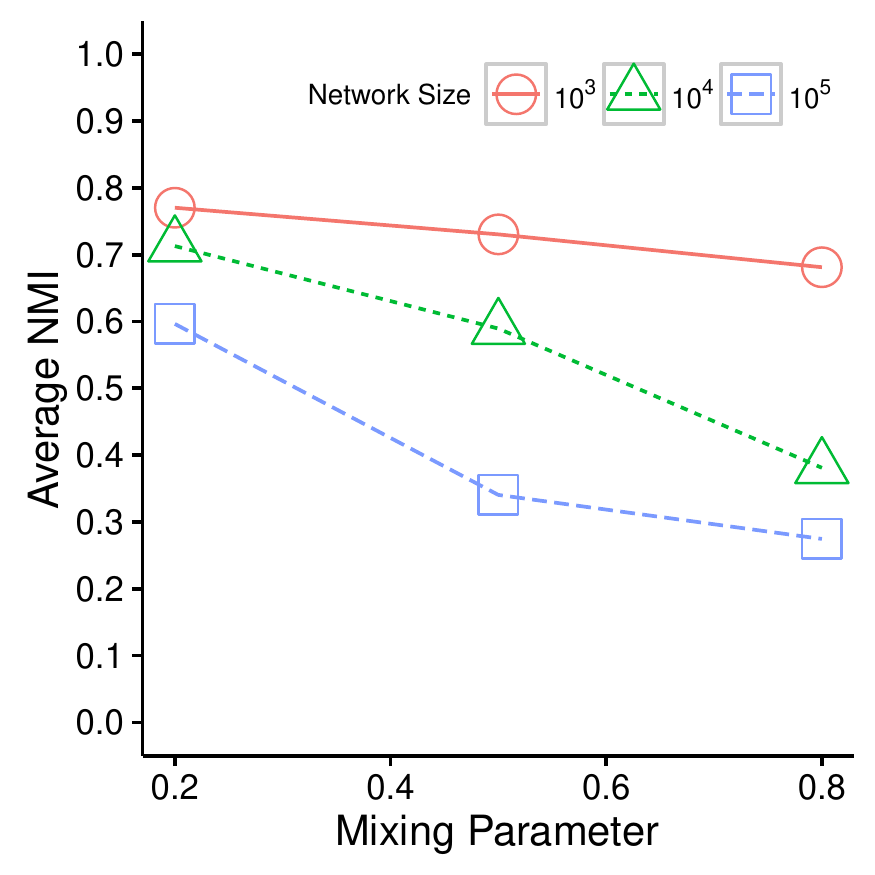}}
\hfill
\subfigure[$\varsigma = 10-15\%$]{\centering \includegraphics[width=0.16\textwidth]{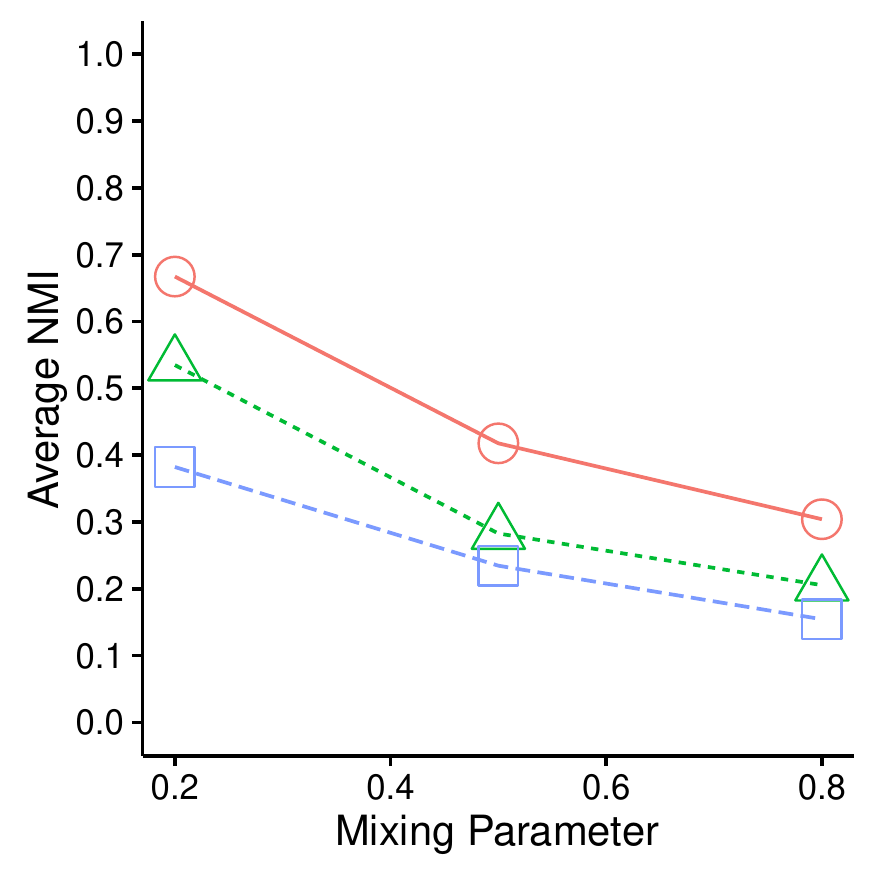}}
\hfill
\subfigure[$\varsigma = 20-30\%$]{\centering \includegraphics[width=0.16\textwidth]{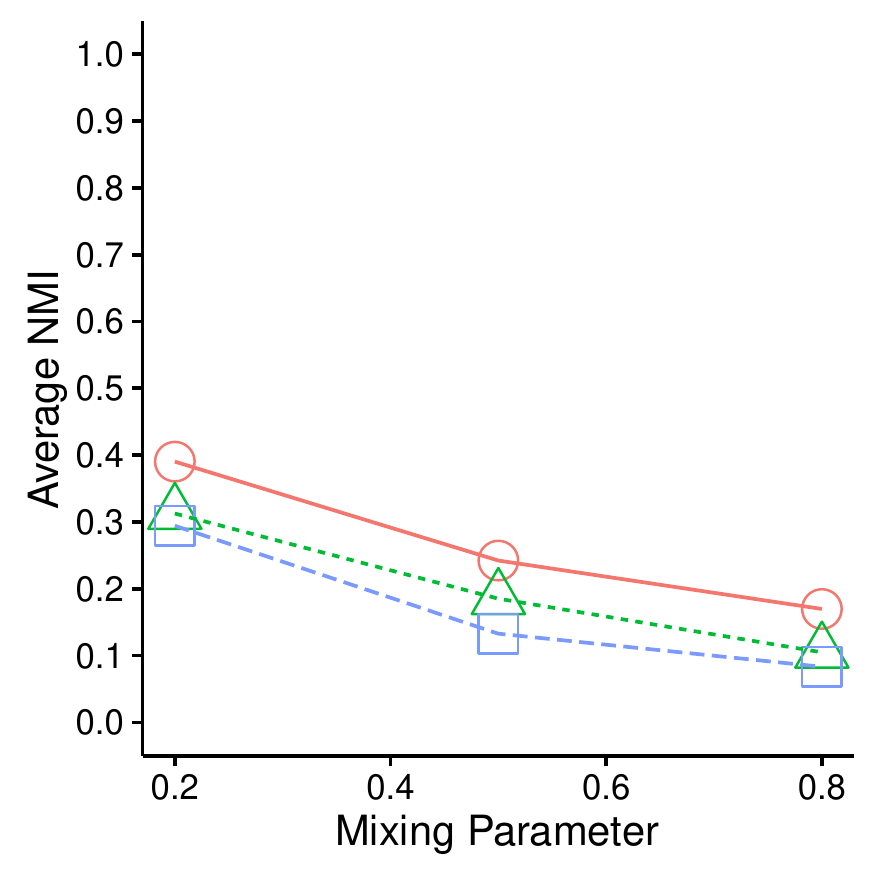}}
\caption{Results for LFR (a,b,c) and NSC (d,e,f) models. The average normalized mutual information as a function of mixing parameter ($\mu =(0.2,0.5,0.8) $). (a,d) $\varsigma=5-10\%$ (b,e)$\varsigma=10-15\%$ (c,f)$\varsigma=20-30\%$  and different network sizes $n=(10^3,10^4,10^5)$ represented by different line markers. The algorithms perform poorly for networks with increasing mixing parameter and for large network (square line marker).}
\label{fig::mixing}
\end{figure}

\section{Results and Discussion}\label{sec::results}

We summarize the results and our findings in this section. We applied different clustering algorithms on the generated benchmark graphs and calculated the Normalized Mutual Information (NMI)(y-axis on all plotted graphs).

Figure \ref{fig::size} presents the results of the two most important topological features for clustering algorithms, the size of networks and the number of clusters. The summarized results for all the algorithms show a decreasing trend in the NMI values as the size of the network is increased (see Figure \ref{fig::size}(a-f) along x-axis). Also, it can be clearly observed that all the clustering algorithms on average, perform poorly when there are a few large sized clusters (see Figure \ref{fig::size}(a-f), represented by round markers) in networks as compared to many small sized clusters (see Figure \ref{fig::size}(a-f), represented by square markers). All the algorithms considered in this study do not require as input, the number of clusters to be generated but rather optimize some objective function to determine when to stop the execution of an algorithm, or they use some dynamic process to determine community structures, irrespective of the cluster sizes. The results clearly reflect the poor performance of all the applied clustering algorithms (see \ref{results} for individual results of all algorithms). These two findings are the most important results of our experimentation, we have shown that the size of networks and the size of clusters have a huge impact on the performance of clustering algorithms. 

The algorithms generally fail to perform well as the size of the networks increases. This directly implies the lack of robustness of clustering algorithms to cope with the explosive increase in the size of networks. This limitation requires us to revisit the optimization functions used to detect and optimize clusters. As new networks with millions and billions of nodes are made available, they present new and interesting challenges for the community to address.

Failing to detect a few large sized clusters but successfully finding many small sized clusters suggests that the clustering algorithms we have studied fail to provide a good summary of the global structure of a network but successfully capture fine details of its topological structure. One alternative solution is the use of hierarchical clustering algorithms rather than just producing partitional clusters. Clustering can be highly subjective and the level of fine details can vary from one problem to another. These details can be the number of clusters, the size of clusters, the ratio of connectivity among nodes and other topological features. The comparative analysis of the ground truth to the found clusters does not suggest that the algorithms are incorrect, but simply points to the often ignored fact that these algorithms somehow capture a detailed view rather than a global view. Thus we can conclude that the design of existing community detection algorithms need to be revised to incorporate these varying structural properties of communities.

Figure \ref{fig::cluster} shows the varying behavior of clustering algorithms with increasing average degree $\langle k \rangle$ of the networks. The results are calculated for different network sizes and cluster sizes. The most interesting result is the behavior of algorithms when $\langle k \rangle$ is low while keeping all other parameters constant. For every plot, we can see an increase in the clustering quality as $\langle k \rangle$ increases. This is due to the fact that most of the clustering algorithms try to find densely connected groups of nodes to be identified as communities. For networks which have low average node connectivity, the clustering algorithms on average perform poorly. 

For Figure \ref{fig::mixing}, the first and previously well known result is the behavior of mixing parameter for all the different graphs. It can be clearly observed that all the algorithms perform poorly for high values of mixing parameter (Fig. \ref{fig::mixing}) irrespective of other parameters. Algorithms perform very well with low values of mixing parameter but for the same value of mixing parameter, when small sized networks are compared to large networks, the performance of algorithms on small networks is much better than on large networks. This re-iterates the earlier discussed results that clustering algorithms perform poorly for large sized networks as compared to networks with small sizes while keeping all other topological features constant.

\section{Conclusions}\label{sec::conclusions}

In this paper, we have studied the performance of network clustering algorithms using two benchmark models. We analyze the collective performance of clustering algorithms and empirically prove several limitations imposed by the topology of networks on the performance of clustering algorithms. This is an important contribution as most of the studies apply these algorithms disregarding the variations in the topological and structural properties of networks and rely on the produced results, which to some extent might not be correct.

As part of the future work, we intend to analyze the individual clustering algorithms in an attempt to overcome their limitations by suggesting different ways to implement them. For example, one possible approach which is currently understudy is the repeated application of algorithms on obtained results to produce clusters at different levels of granularity. Other possible improvements include revisiting objective functions and dynamic processes which could lead to better clustering results.

\appendix
\section*{APPENDIX}
\setcounter{section}{1}

The appendix contains two subsections. The first section presents the detailed results for all the clustering algorithms applied on all the benchmark networks generated using the two synthetic network generation models. The second section of appendix contains basic network metrics calculated for networks generated using the newly proposed NSC model. These metrics are provided to demonstrate that the networks generated using this model indeed have small world and scale free properties. These properties  provide a good guideline to prove that the generated networks are similar to real world networks.


\appendixhead{ZHOU}

\begin{acks}
The authors would like to thank Lora Webb of MetricAid for important suggestions to improve the manuscript.
\end{acks}

\bibliographystyle{ACM-Reference-Format-Journals}
\bibliography{visu}

\received{June 2016}{June 2016}{June 2016}

\elecappendix

\medskip

\section{Detailed Results of Clustering Algorithms}\label{results}

This section presents the details of the results obtained for individual clustering algorithms for all the different parameter variations used to evaluate the community detection algorithms.

\subsection{Studying the Effects of Network Size}\label{sec::size}

\begin{figure}
\centering
\subfigure[$\langle k \rangle$=3]{\centering \includegraphics[width=0.32\textwidth]{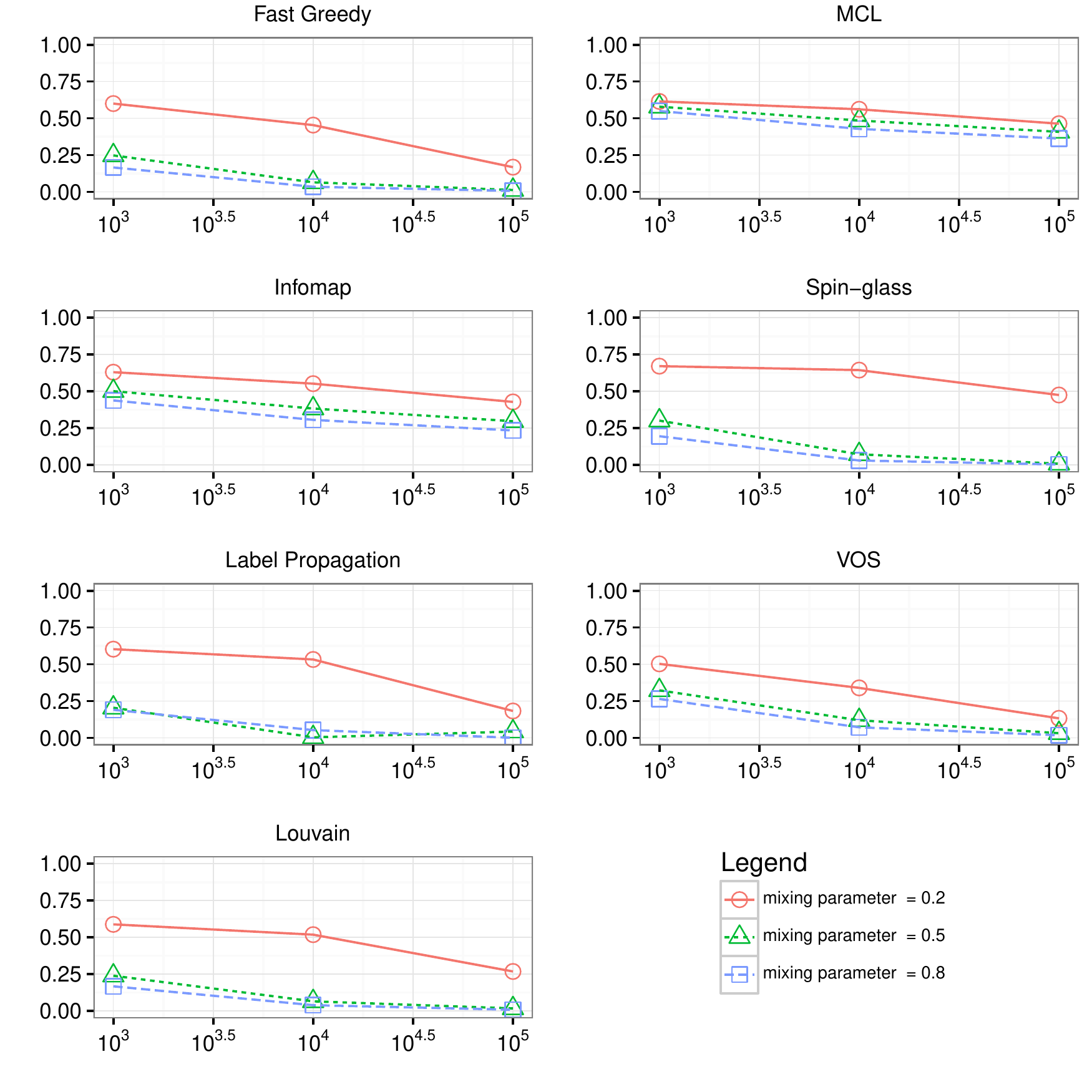} \label{fig::sizeA}}
\hfill
\subfigure[$\langle k \rangle$=5]{\centering \includegraphics[width=0.32\textwidth]{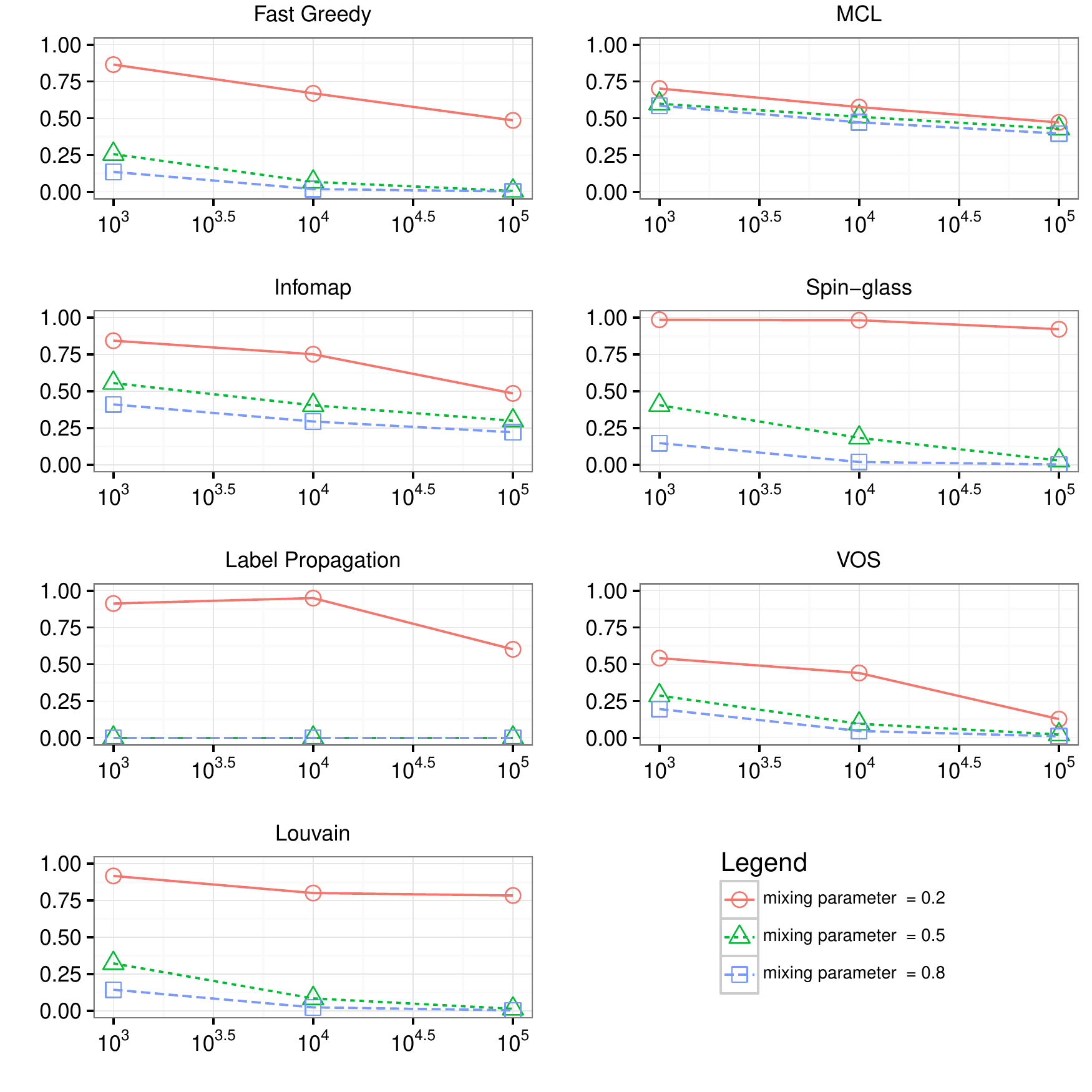}\label{fig::sizeB}}
\hfill
\subfigure[$\langle k \rangle$=10]{\centering \includegraphics[width=0.32\textwidth]{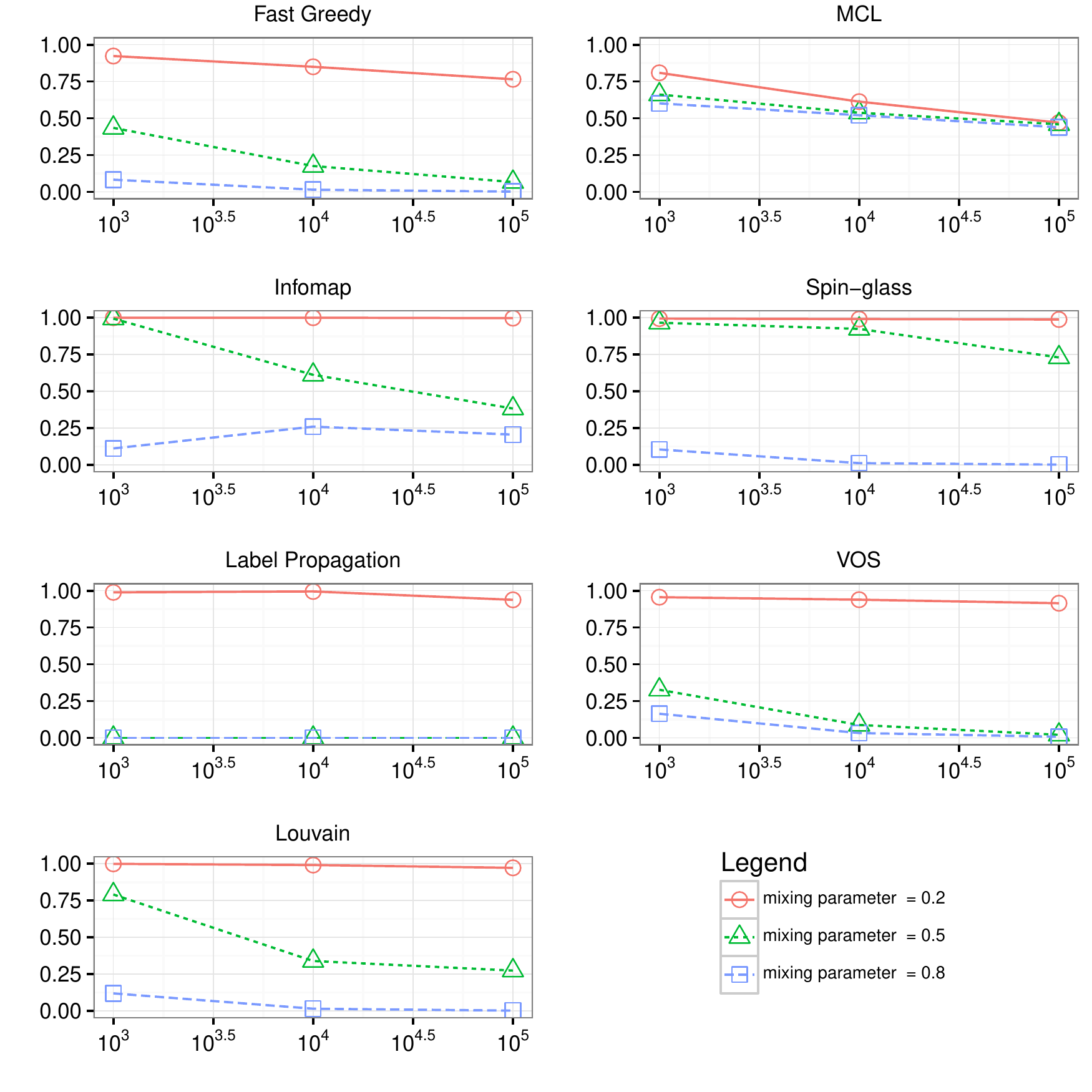}\label{fig::sizeC}}
\textbf{Results for LFR Benchmark: Increasing Network Size}\par\medskip

\subfigure[$\langle k \rangle$=3]{\centering \includegraphics[width=0.32\textwidth]{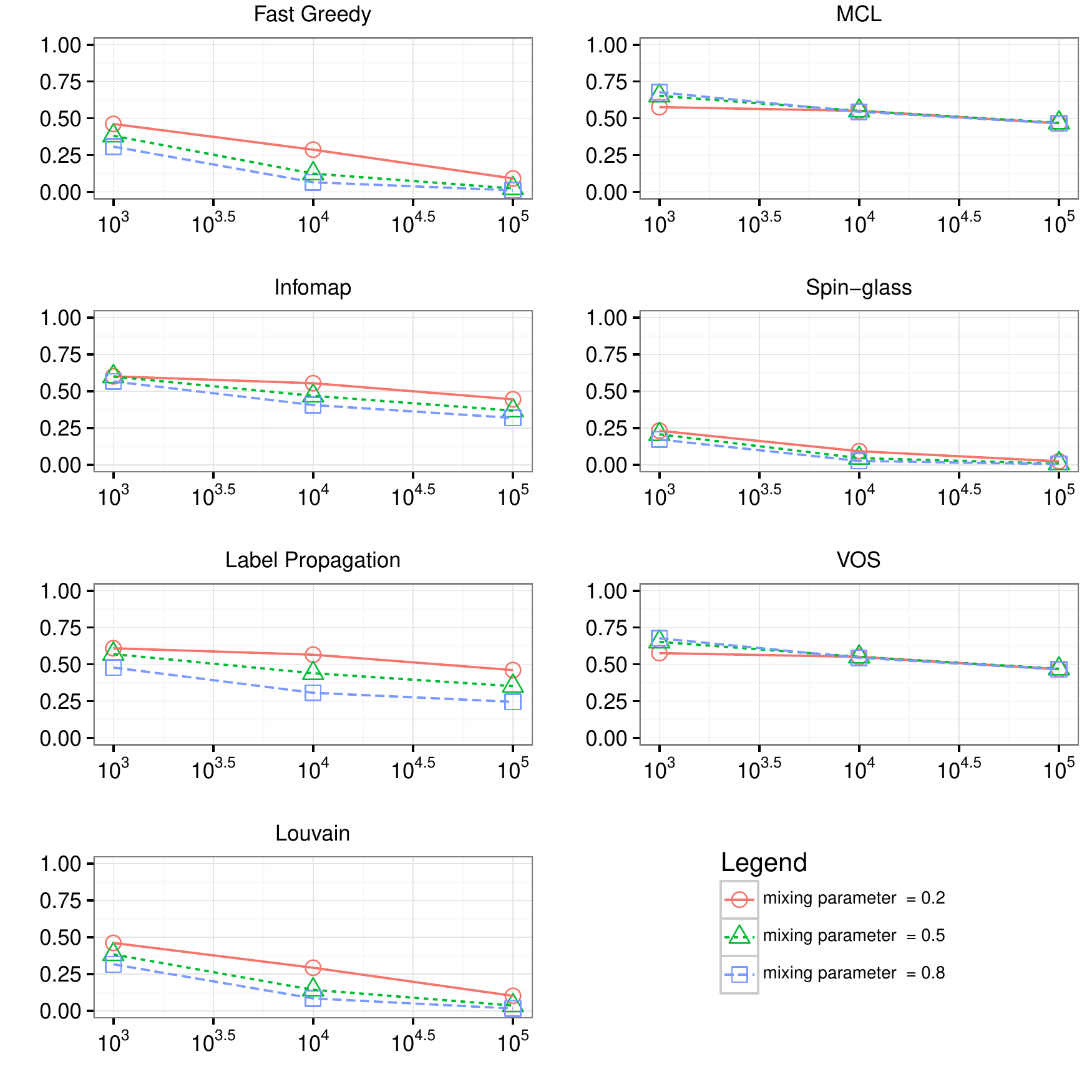}\label{fig::sizeD}}
\hfill
\subfigure[$\langle k \rangle$=5]{\centering \includegraphics[width=0.32\textwidth]{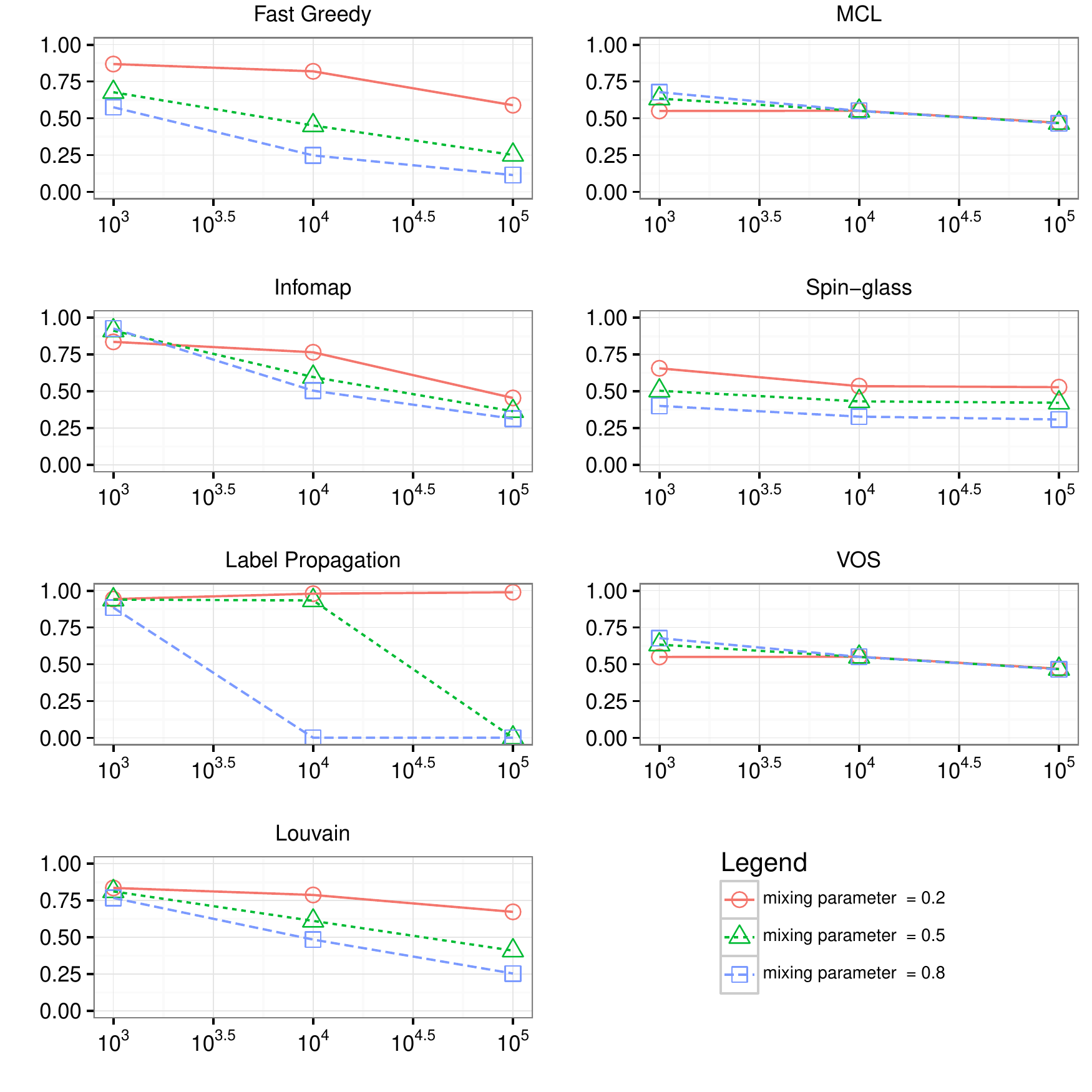}\label{fig::sizeE}}
\hfill
\subfigure[$\langle k \rangle$=10]{\centering \includegraphics[width=0.32\textwidth]{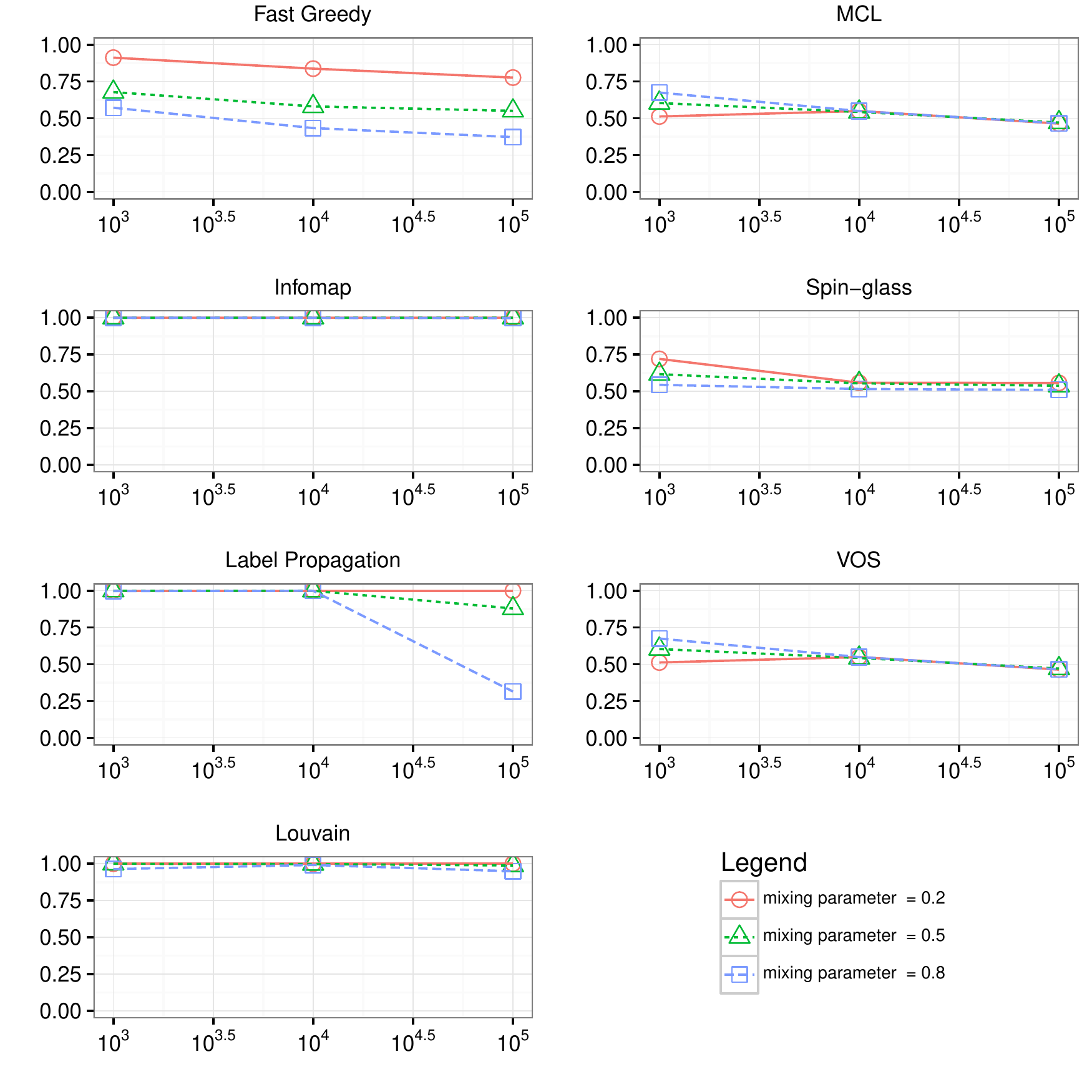}\label{fig::sizeF}}
\textbf{Results for NSC Benchmark: Increasing Network Size}\par\medskip
\caption{Results for LFR (a,b,c) amd NSC (d,e,f) models. The average normalized mutual information as a function of network size($n$) with $n=(10^3,10^4,10^5)$. (a,d) Mixing Parameter $\mu=0.2$, (b,e) $\mu=0.5$ and (c,f) $\mu=0.8$. The graphs present the individual results of all the seven clustering algorithms. Algorithms generally perform poorly with increasing size of networks with respect to other networks with similar topological features.}%
\label{fig::sizeAppendix}
\end{figure}

Figure \ref{fig::size} of the article summarized the results of different clustering algorithms as a function of increasing network sizes in terms of number of nodes. Figure \ref{fig::sizeAppendix} provides the details of these results by presenting all the individual results for each clustering algorithm and the varying parameters. The results obtained from LFR model are presented in Figure \ref{fig::sizeA},\ref{fig::sizeB},\ref{fig::sizeC} and that of NSC model are presented in Figure \ref{fig::sizeD},\ref{fig::sizeE},\ref{fig::sizeF}.

The subfigures in Figure \ref{fig::sizeAppendix} correspond to the results compiled for the two models and the seven clustering algorithms used for clustering. The results are presented for three different values of average node connectivity $(\langle k \rangle=\{3, 5, 10\})$. The increasing network sizes are plotted along the x-axis $(nodes=\{10^3, 10^4, 10^5\})$ and NMI values are plotted on the y-axis. The variations in mixing parameter $(\mu=\{0.2, 0.5, 0.8\})$ are represented by different line markers $(circle, triagle,square)$.



%
As the number of nodes are increased, all the algorithms generally decline in the quality of clusters obtained. One slight exception to this behavior is the InfoMap algorithm  for LFR benchmark when the $\langle k \rangle$=10 as there is a very small increase in the NMI value when the size of the network is changed from $10^3$ to $10^4$, but the NMI value drops again as the size is further increased to $10^5$. Apart from this negligible increase, all the algorithms show a decrease in the clustering quality as the network size is increased. This clearly shows that as network sizes increase, clustering algorithms have some difficulty finding community structures.


\subsection{Studying the Effects of Cluster Sizes}\label{sec::cluster}

\begin{figure}
\centering

\subfigure[$\mu = 0.2$]{\centering \includegraphics[width=0.32\textwidth]{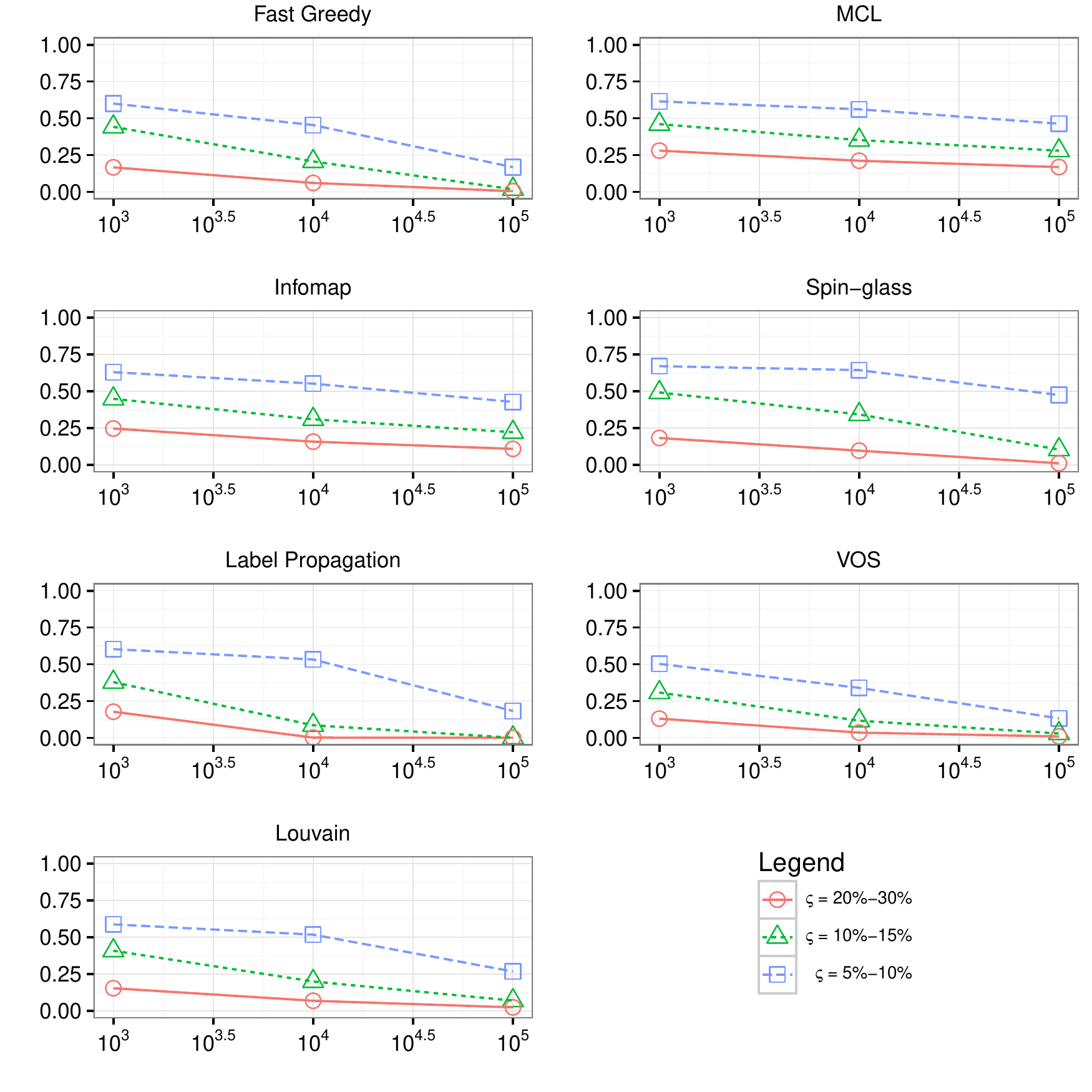}\label{fig::clusterA}}
\hfill
\subfigure[$\mu = 0.5$]{\centering \includegraphics[width=0.32\textwidth]{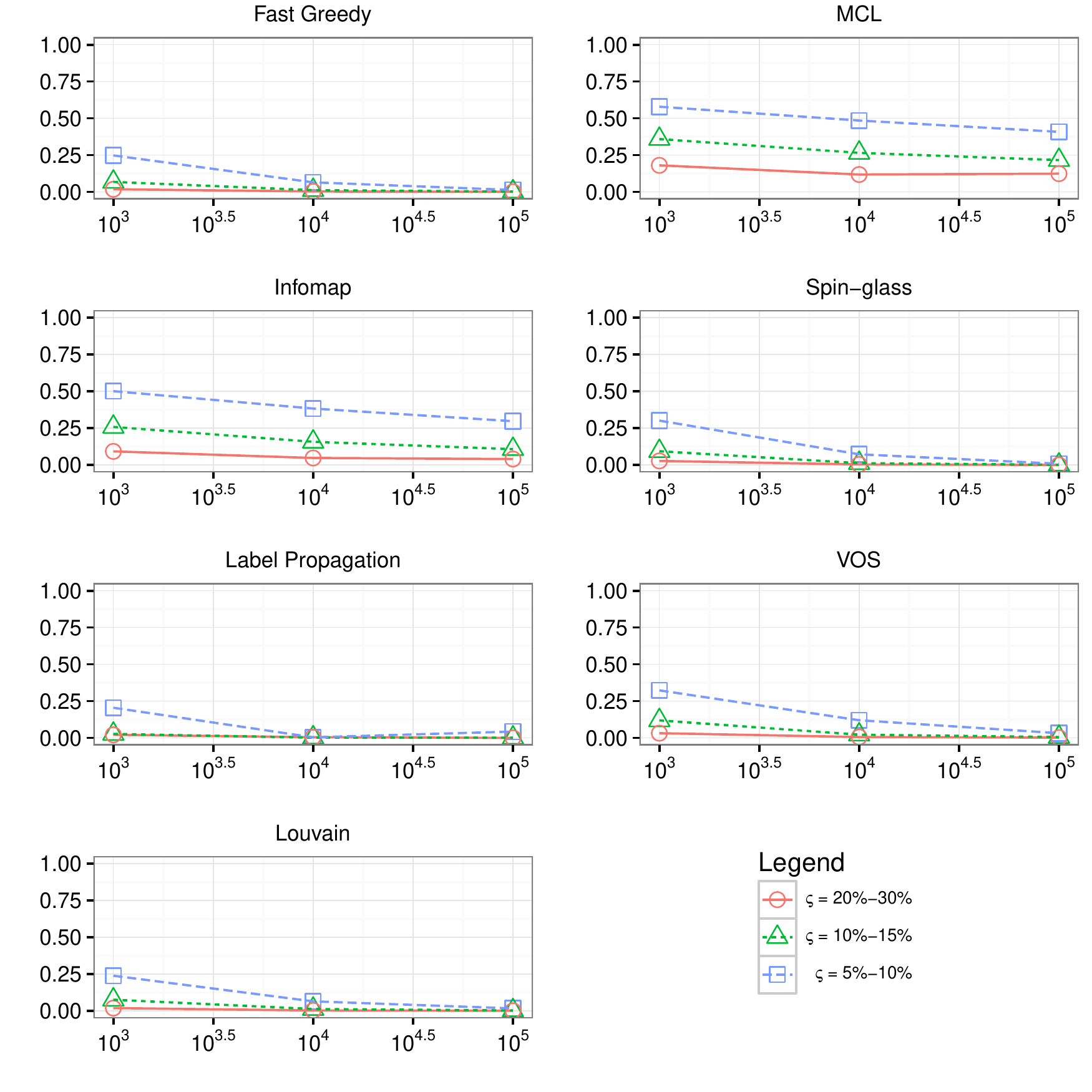}\label{fig::clusterB}}
\hfill
\subfigure[$\mu = 0.8$]{\centering \includegraphics[width=0.32\textwidth]{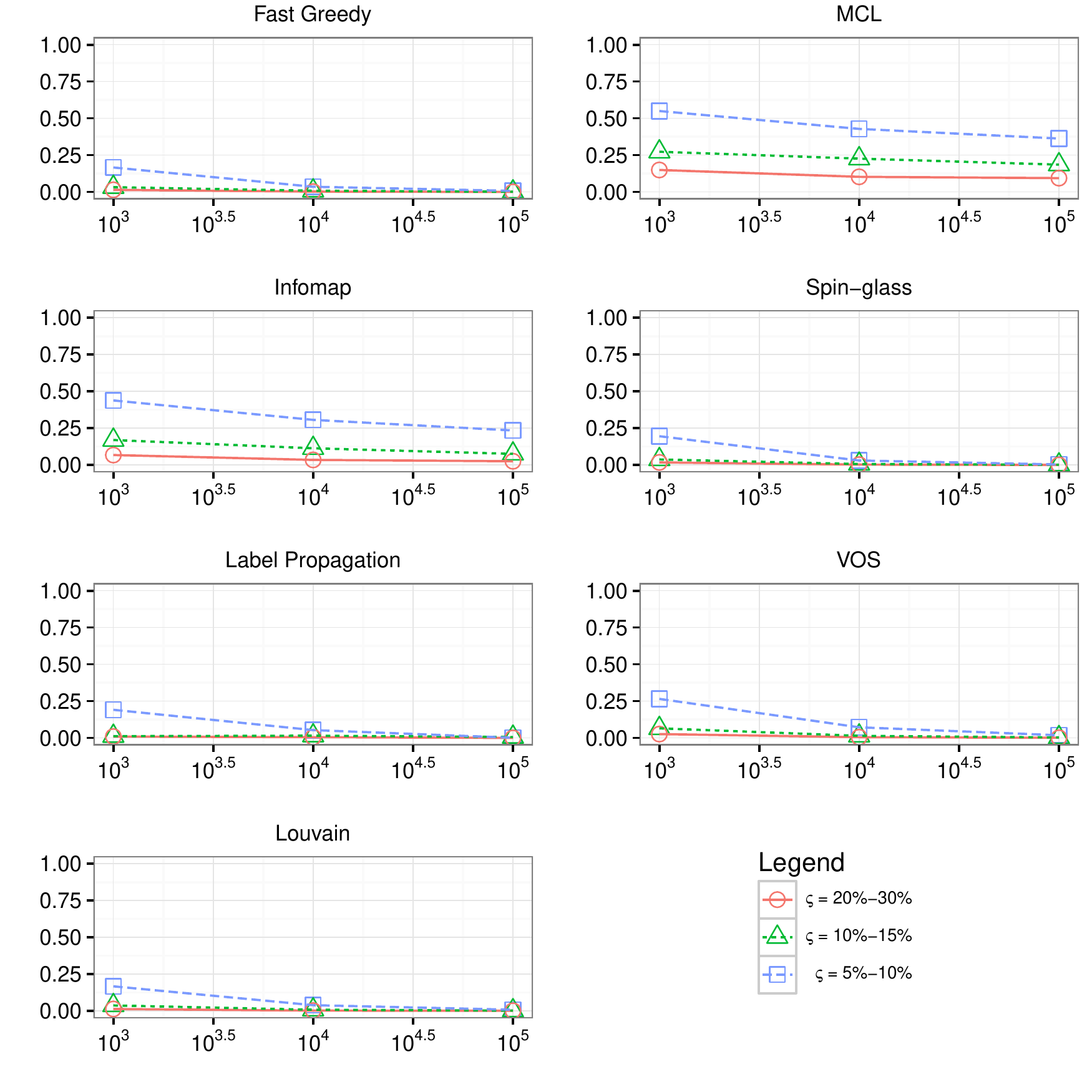}\label{fig::clusterC}}
\textbf{Results for LFR Benchmark: Changing Cluster Sizes}\par\medskip
\subfigure[$\mu = 0.2$]{\centering \includegraphics[width=0.32\textwidth]{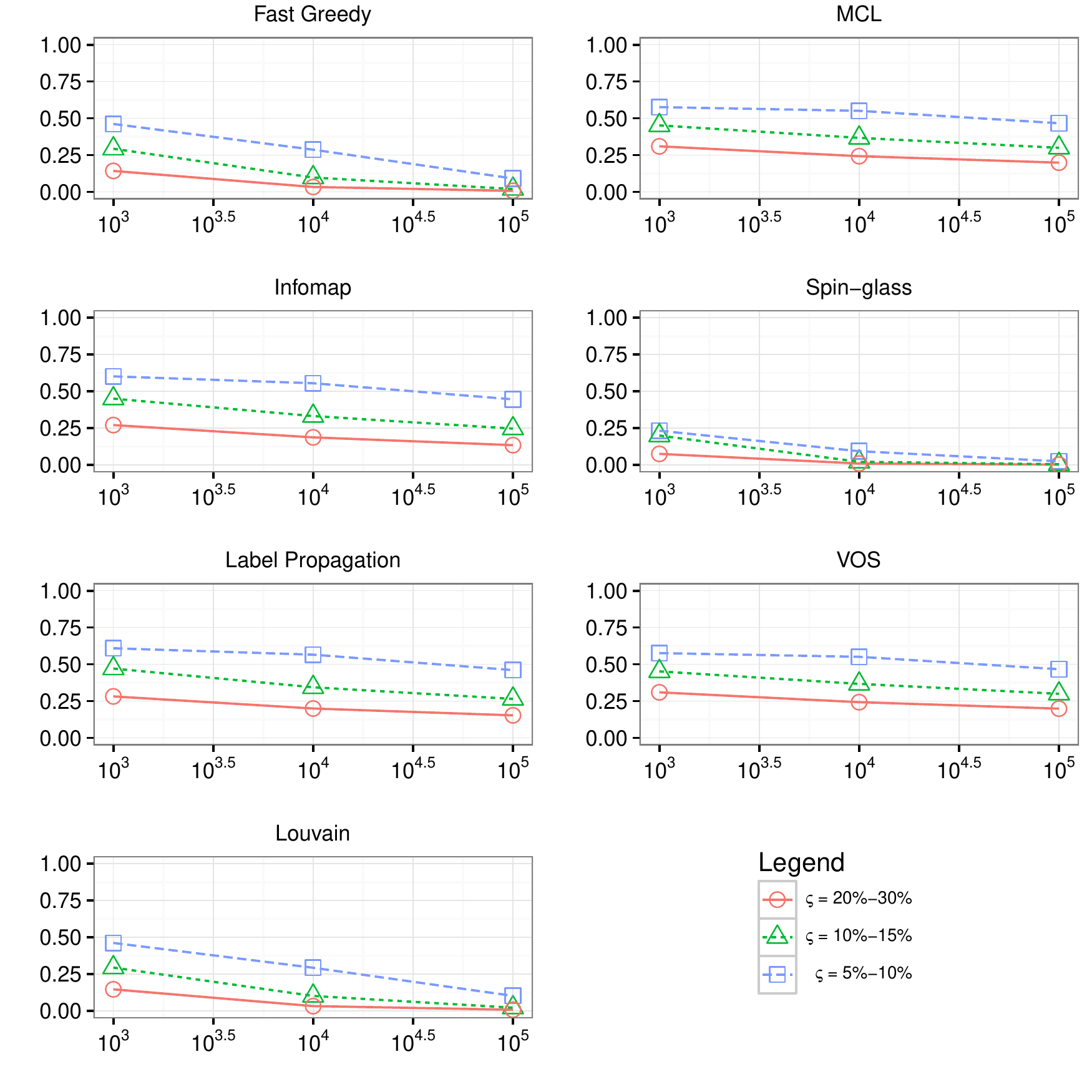}
\label{fig::clusterD}}
\hfill
\subfigure[$\mu = 0.5$]{\centering \includegraphics[width=0.32\textwidth]{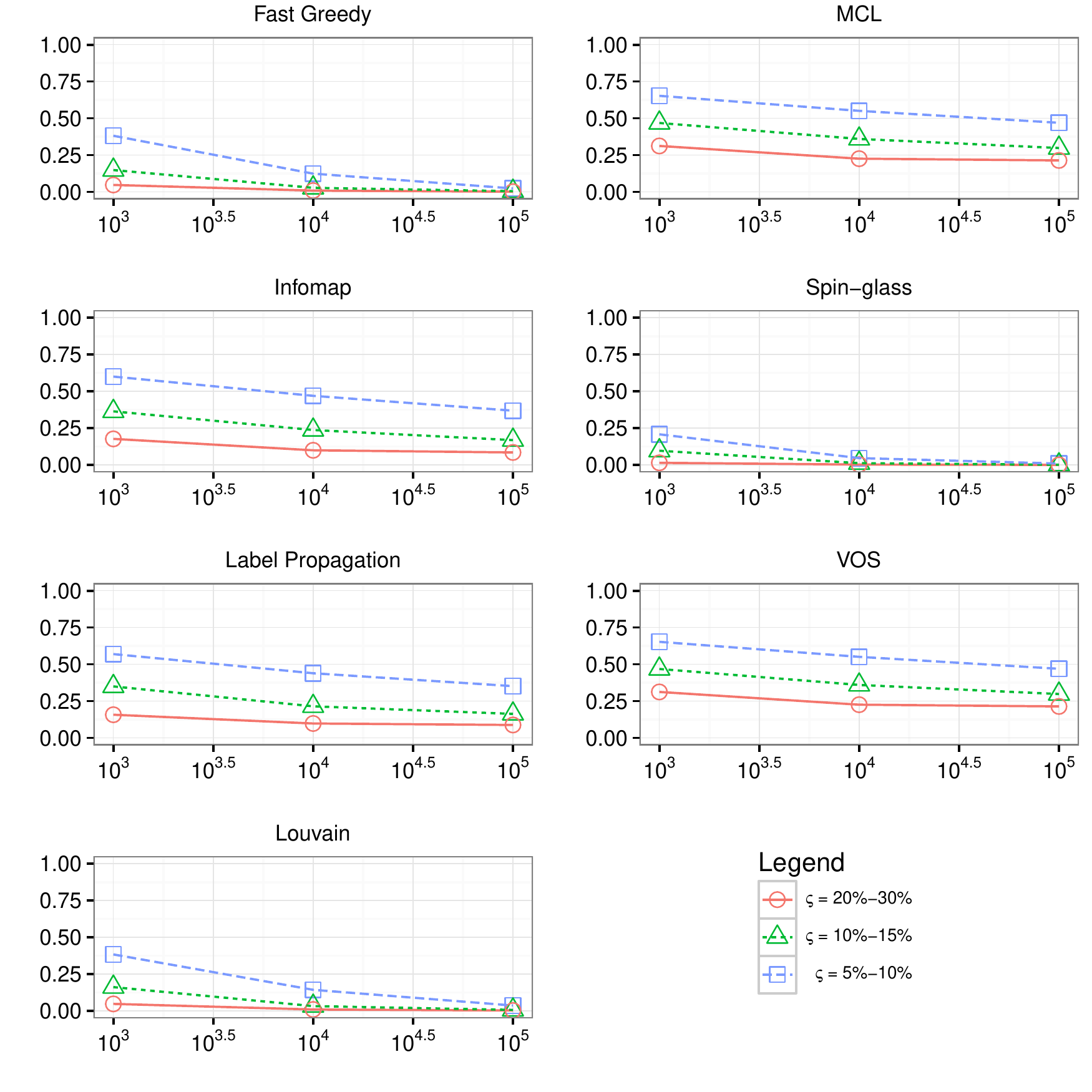}\label{fig::clusterE}}
\hfill
\subfigure[$\mu = 0.8$]{\centering \includegraphics[width=0.32\textwidth]{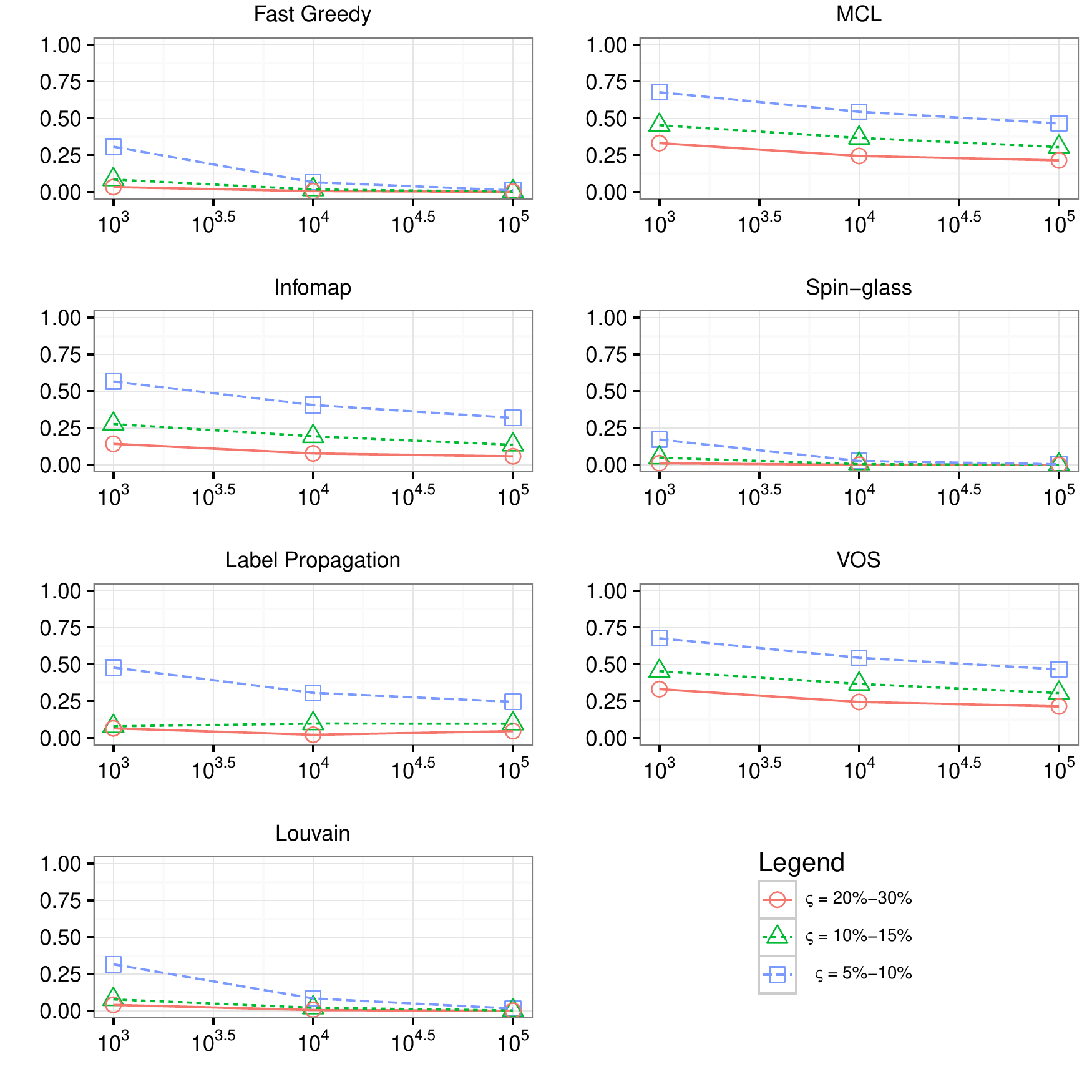}
\label{fig::clusterF}}
\textbf{Results for NSC Benchmark: Changing Cluster Sizes}\par\medskip
\caption{Results for LFR (a,b,c) amd NSC (d,e,f) models. The average normalized mutual information as a function of cluster varying sizes. The graphs present the individual results of all the seven clustering algorithms. Algorithms generally perform poorly when the size of the clusters is very large.}
\label{fig::clusterAppendix}
\end{figure}

Figure \ref{fig::cluster} of the article summarized the results of different clustering algorithms as a function of increasing number of nodes and varying number of communities and their sizes. Figure \ref{fig::clusterAppendix} of provides the details of these results by presenting all the individual results for each clustering algorithm and the varying parameters. The results obtained from LFR model are presented in Figure \ref{fig::clusterA},\ref{fig::clusterB},\ref{fig::clusterC} and that of NSC model are presented in Figure \ref{fig::clusterD},\ref{fig::clusterE},\ref{fig::clusterF}.

The subfigures in Figure \ref{fig::clusterAppendix} correspond to the results compiled for the two models and the seven clustering algorithms used for clustering. The variations in community sizes are represented by different line markers. The results for plotted for three different values of network sizes $(n=\{10^3,10^4,10^5\})$. 

For each plot, the performance of clustering algorithms is better when there are a few large sized clusters in a network labelled $20\%-30\%$ (which is the minimum to maximum size of clusters in terms of total number of nodes and represented by round line markers). Clustering algorithms perform well when they are required to find many small sized clusters in a network which highlights another important limitation of clustering algorithms.

\subsection{Studying the Effects of Average Degree}\label{sec::average}

\begin{figure}
\centering

\subfigure[$\mu = 0.2$]{\centering \includegraphics[width=0.32\textwidth]{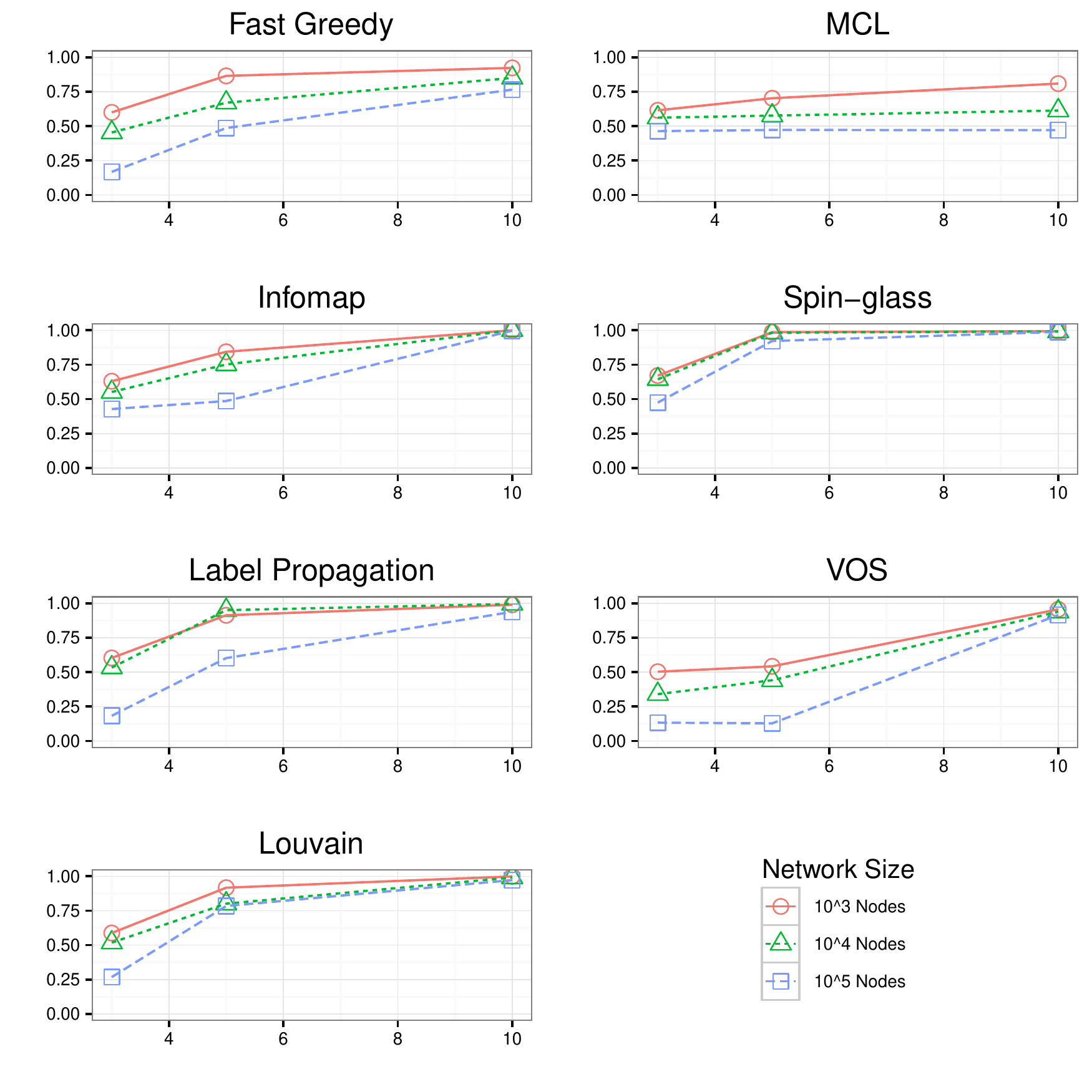}\label{fig::mixingA}}
\hfill
\subfigure[$\mu = 0.5$]{\centering \includegraphics[width=0.32\textwidth]{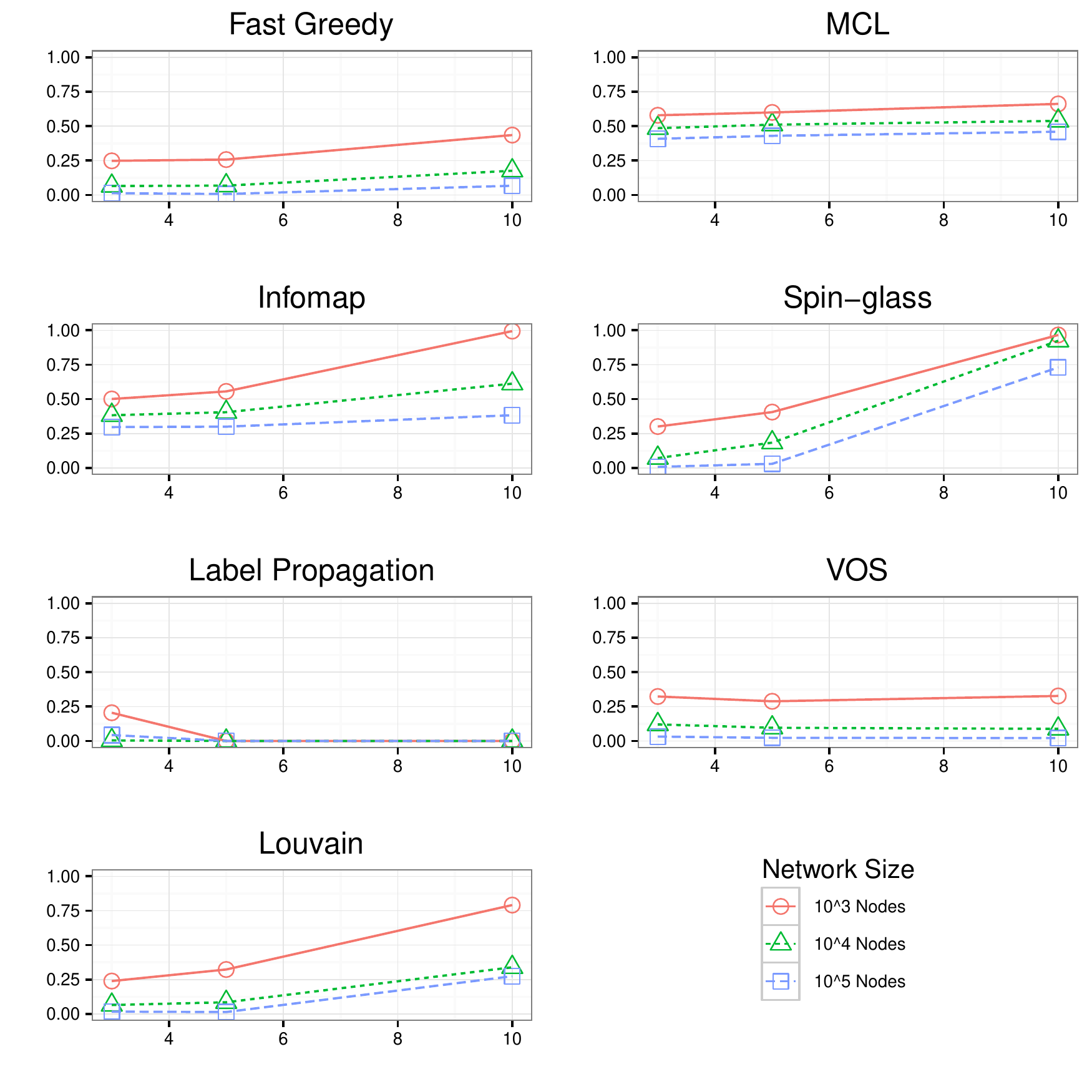}\label{fig::mixingB}}
\hfill
\subfigure[$\mu = 0.8$]{\centering \includegraphics[width=0.32\textwidth]{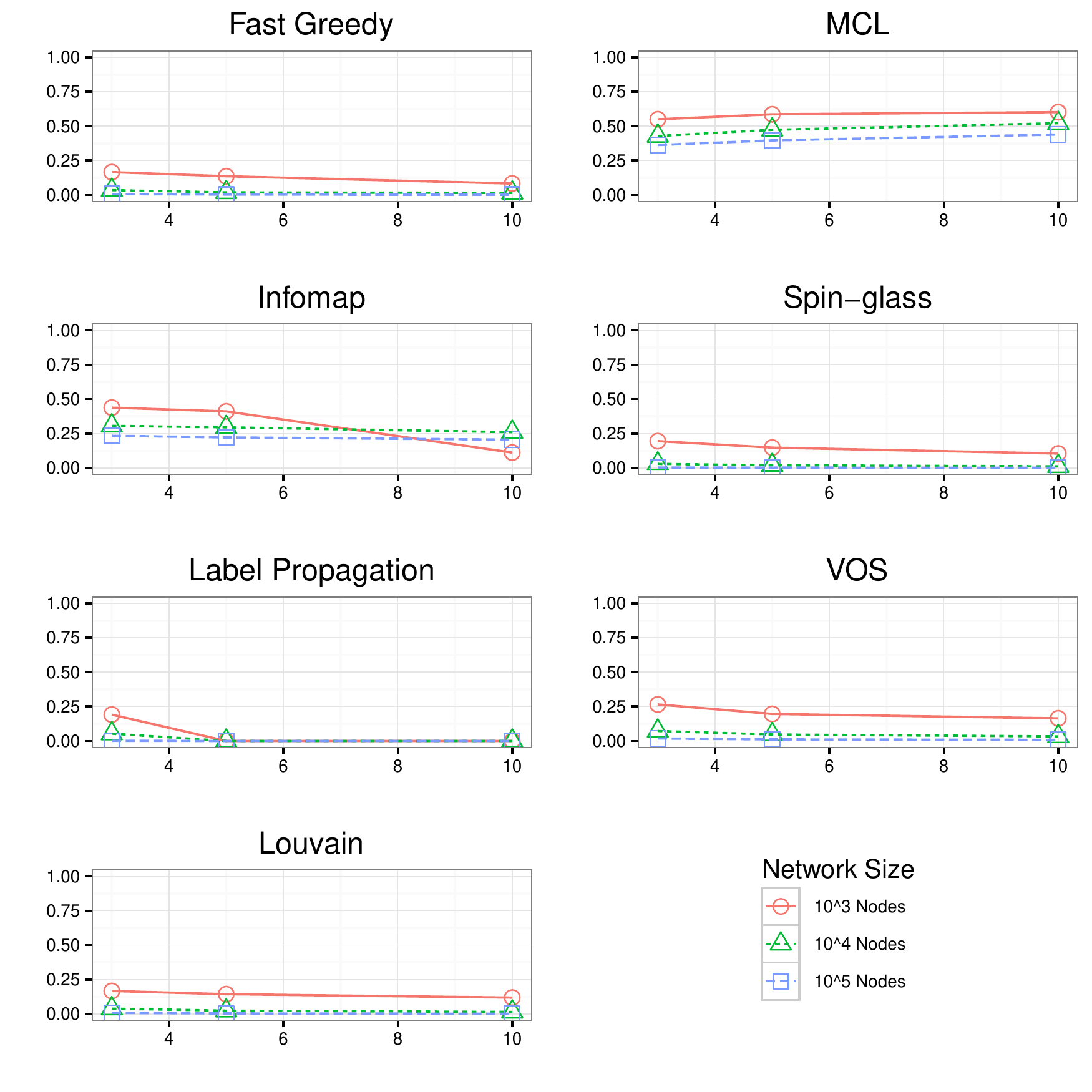}
\label{fig::mixingC}}
\textbf{Results for LFR Benchmark: Increasing Average Degree}\par\medskip

\subfigure[$\mu = 0.2$]{\centering \includegraphics[width=0.32\textwidth]{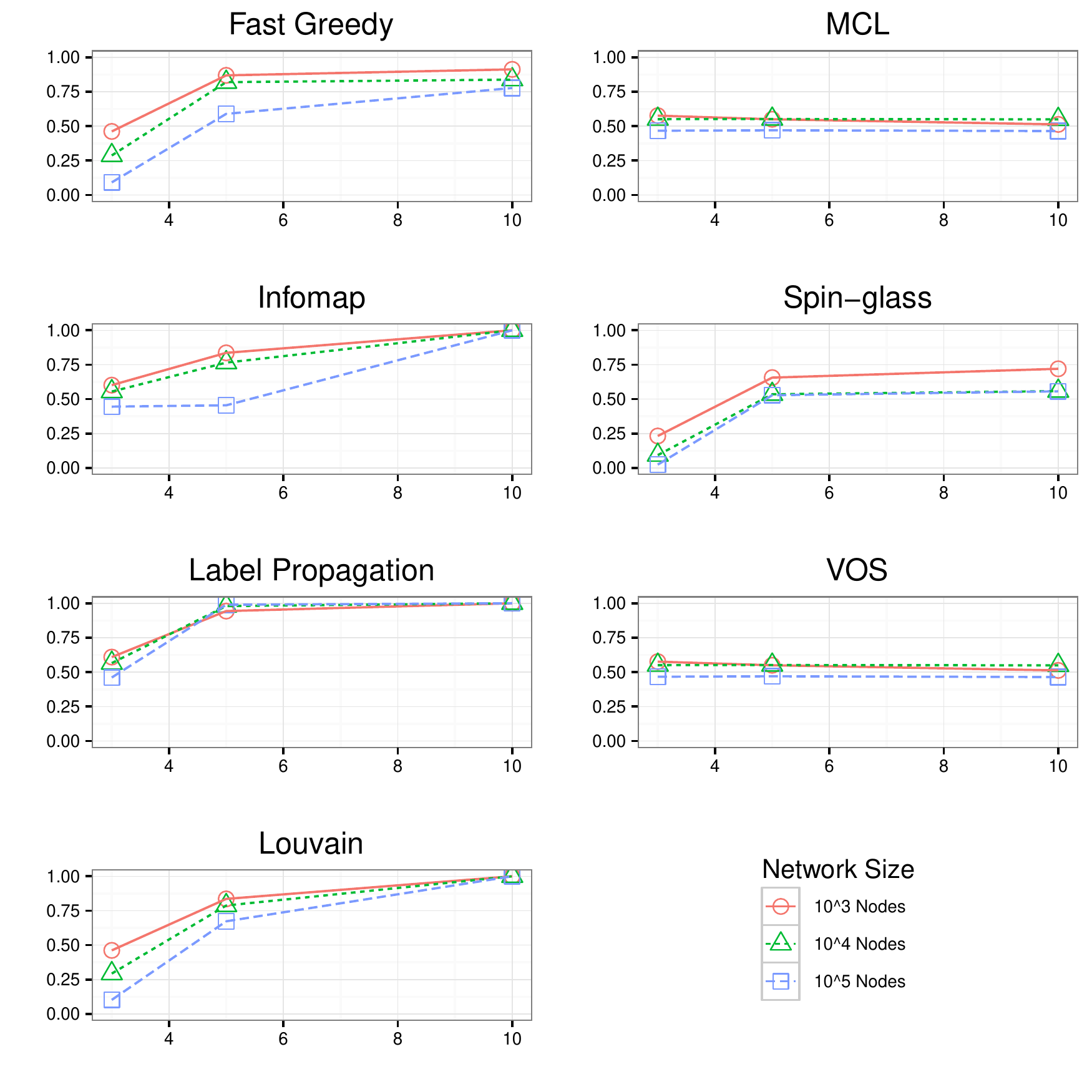}\label{fig::mixingD}}
\hfill
\subfigure[$\mu = 0.5$]{\centering \includegraphics[width=0.32\textwidth]{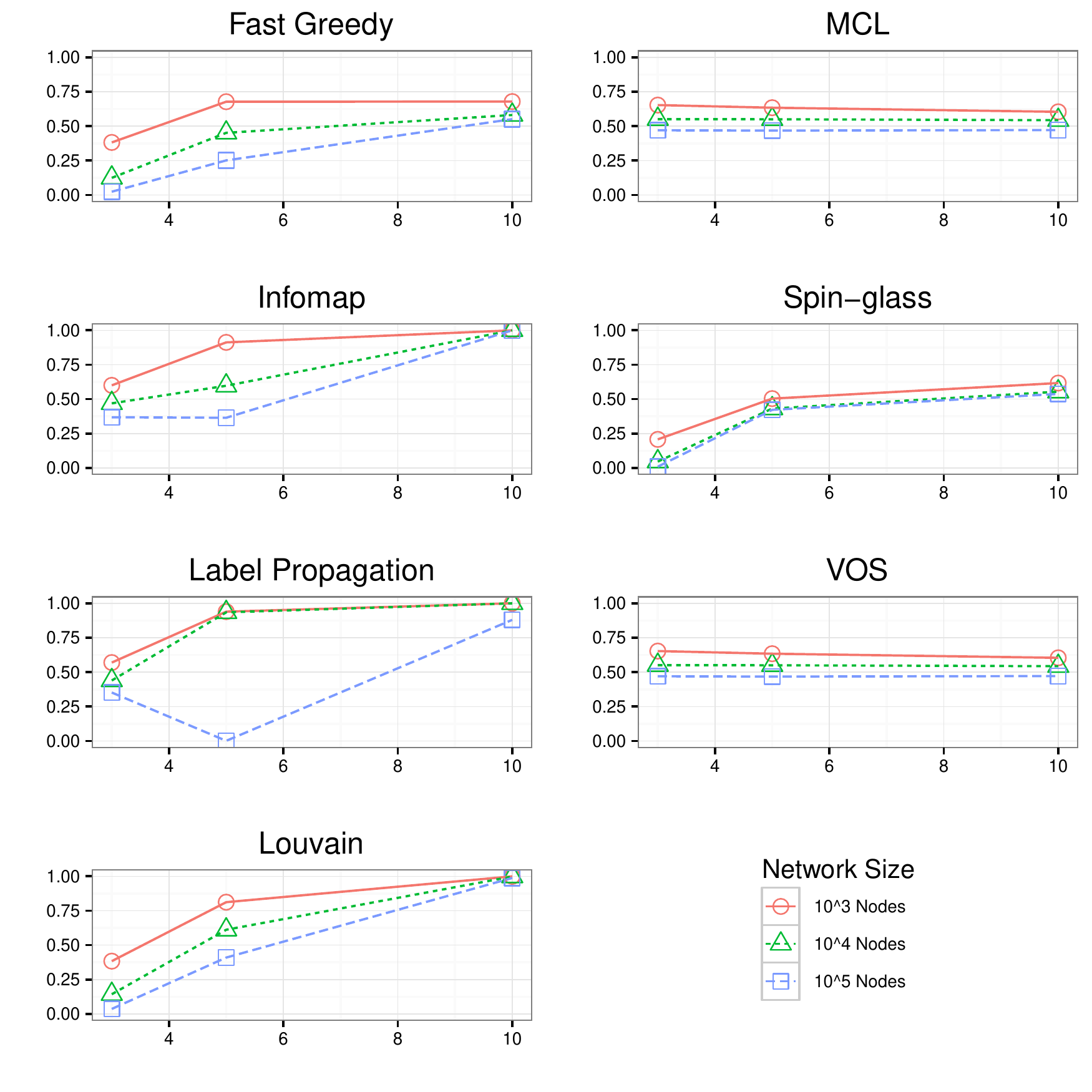}
\label{fig::mixingE}}
\hfill
\subfigure[$\mu = 0.8$]{\centering \includegraphics[width=0.32\textwidth]{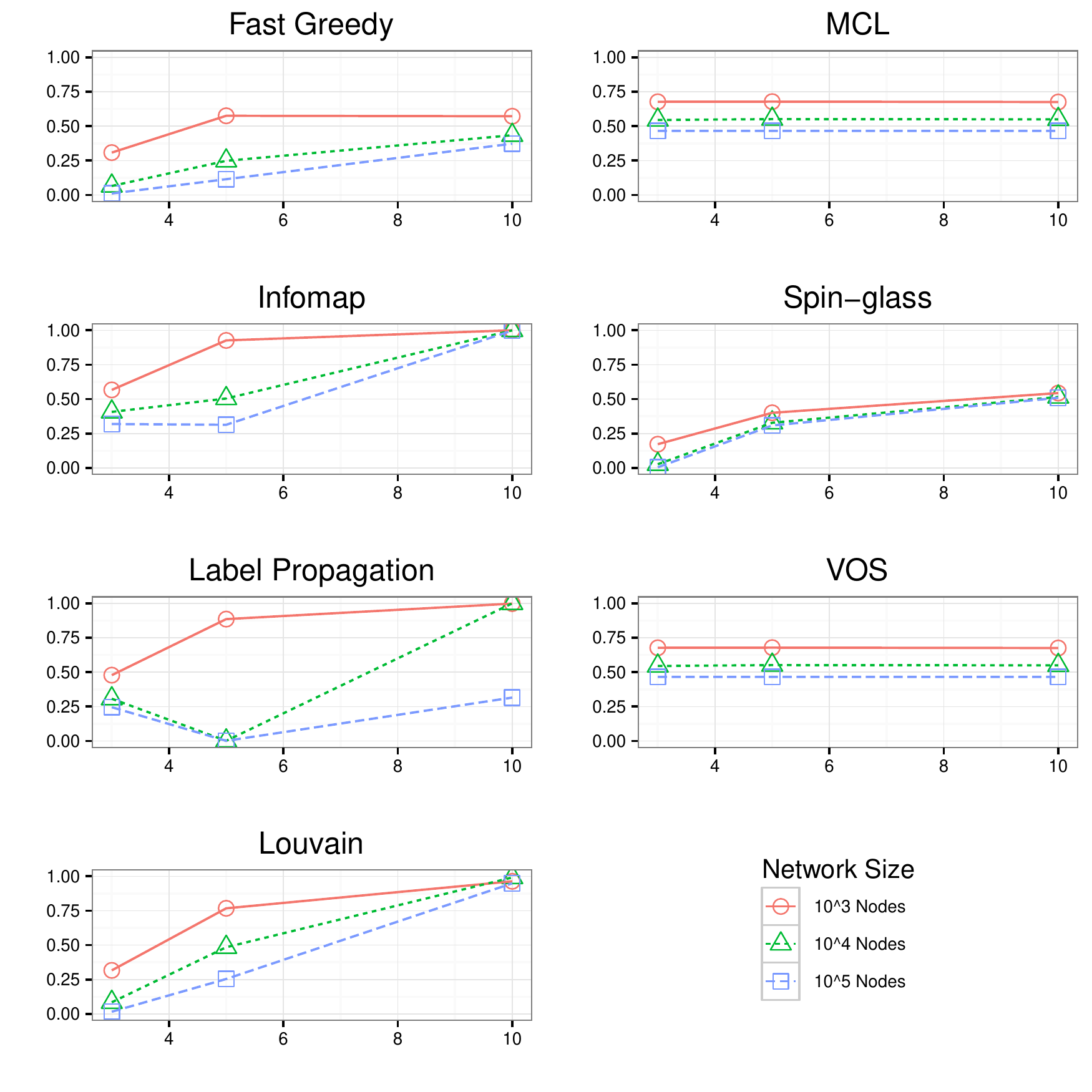}\label{fig::mixingF}}
\textbf{Results for NSC Benchmark: Increasing Average Degree}\par\medskip
\caption{Results for LFR (a,b,c) amd NSC (d,e,f) models. The average normalized mutual information as a function of $(\langle k \rangle=\{3, 5, 10\})$. The graphs present the individual results of all the seven clustering algorithms. Algorithms generally perform better with the increasing average node degree.}%
\label{fig::mixingAppendix}
\end{figure}

Figure \ref{fig::mixing} of the article summarized the results of different clustering algorithms as a function of increasing average degree or connectivity of nodes. Figure \ref{fig::mixingAppendix} provides the details of these results by presenting all the individual results for each clustering algorithm and the varying parameters. The results obtained from LFR model are presented in Figure \ref{fig::mixingA}, \ref{fig::mixingB}, \ref{fig::mixingC} and that of NSC model are presented in Figure \ref{fig::mixingD}, \ref{fig::mixingE}, \ref{fig::mixingF}.

The subfigures in Figure \ref{fig::mixingAppendix} correspond to the results compiled for the two models and the seven clustering algorithms used for clustering. The average degree increases along x-axis $(\langle k \rangle=\{3, 5, 10\})$ while three different network sizes $(nodes=\{10^3, 10^4, 10^5\})$ are represented by different line markers. 

The results demonstrate the behavior of clustering algorithms generally improve as the average degree increases. For both the LFR and the NSC benchmarks, the results clearly improve for $\mu=0.2$. For the LFR benchmark and $\mu=0.5$, the results are again consistent with notable improvements in the clustering quality but with NSC benchmark, Label Propagation is an exception when $\langle k \rangle=5$ and the network size is $10^5$ (see Figure \ref{fig::mixingE}) as NMI decreases from when $\langle k \rangle=3$ to $\langle k \rangle=5$, but increases again for $\langle k \rangle=10$. This strange behavior needs to be further explored with detailed analysis of the clustering algorithm in order to find an explanation. The results for the LFR benchmark when $\mu=0.8$ generally show a negligible change and are very poor. This is due to the high mixing of nodes with nodes of other clusters. Algorithms perform fairly well for the NSC benchmark even when $\mu=0.8$, this is due to the inherent structure of the networks and the way they are generated. The only exception is again for Label Propagation when  $\langle k \rangle=5$ and network sizes are $10^4, 10^5$. For all the remaining cases, the results improve with an increase in the average degree of the networks.

%

\section{Naive Scale-free Clustering}\label{sec::implementation}

\begin{figure}[!]
	\centering
	\subfigure[$\mu = 0.2$]{\centering \includegraphics[width=0.24\textwidth]{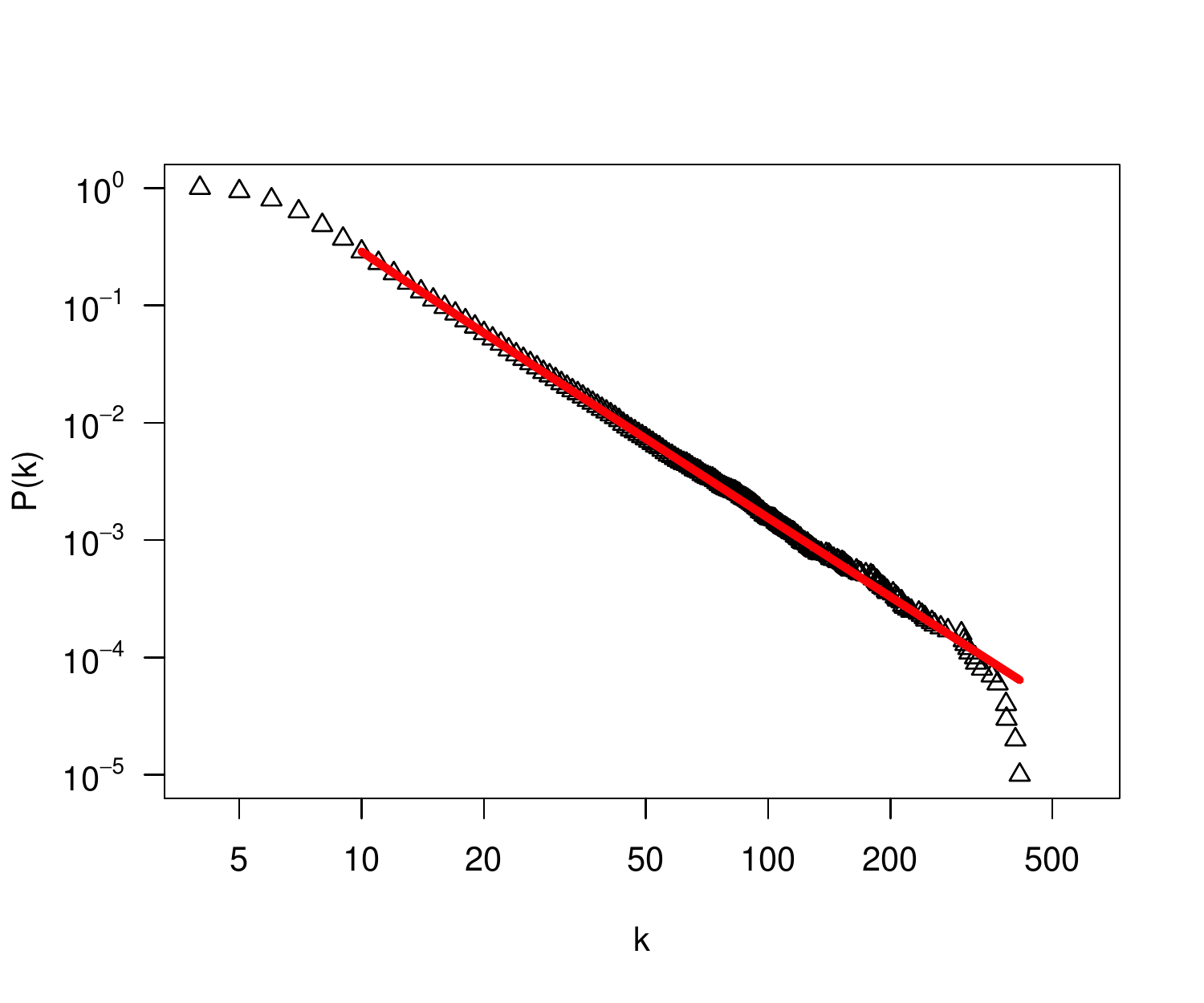}\label{fig::degreeA}}
	\hfill
	\subfigure[$\mu = 0.5$]{\centering \includegraphics[width=0.24\textwidth]{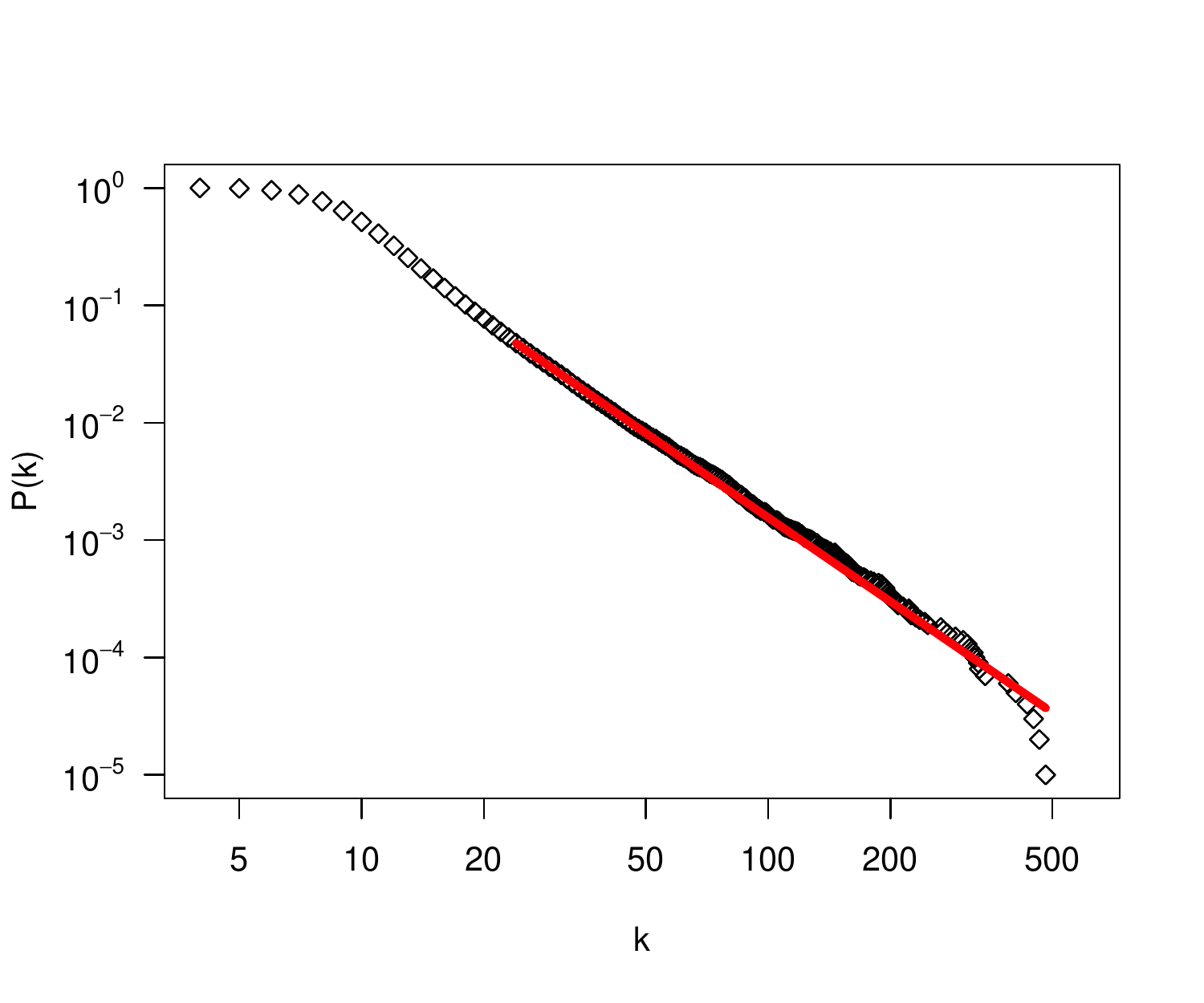}\label{fig::degreeB}}
	\hfill
	\subfigure[$\mu = 0.8$]{\centering \includegraphics[width=0.24\textwidth]{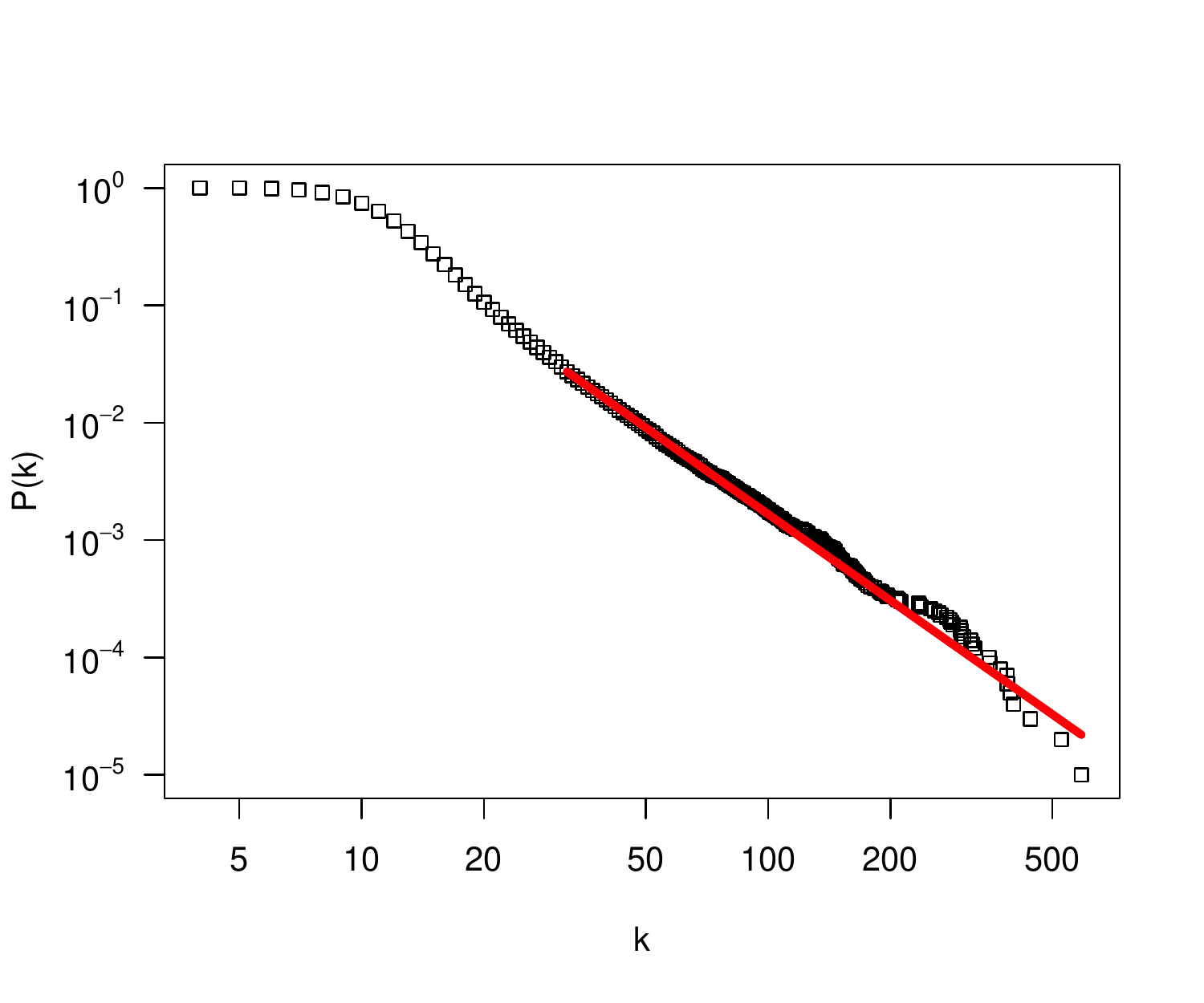}\label{fig::degreeC}}
	\caption{The logarithmic degree distribution of nodes in networks generated using NSC model. The size of the networks is $10^5$ nodes with an average degree of $10$. The power-law fit $\gamma$ is \ref{fig::degreeA} $-3.24$, \ref{fig::degreeB} $-3.7$ and \ref{fig::degreeC} $-3.43$ }
	\label{fig::degree}
\end{figure}

We also studied the topological properties of the newly proposed NSC model to generate benchmark networks. Degree distribution of all the generated graphs follows power law with exponent $ \gamma = (2,4)$ which is consistent with the original BA model \cite{barabasi99}. Figure \ref{fig::degree} shows the degree distribution of graphs generated by the model of size $10^5$ for different values of mixing parameter. 

\begin{figure}[b!]
	\includegraphics[width=0.8\textwidth]{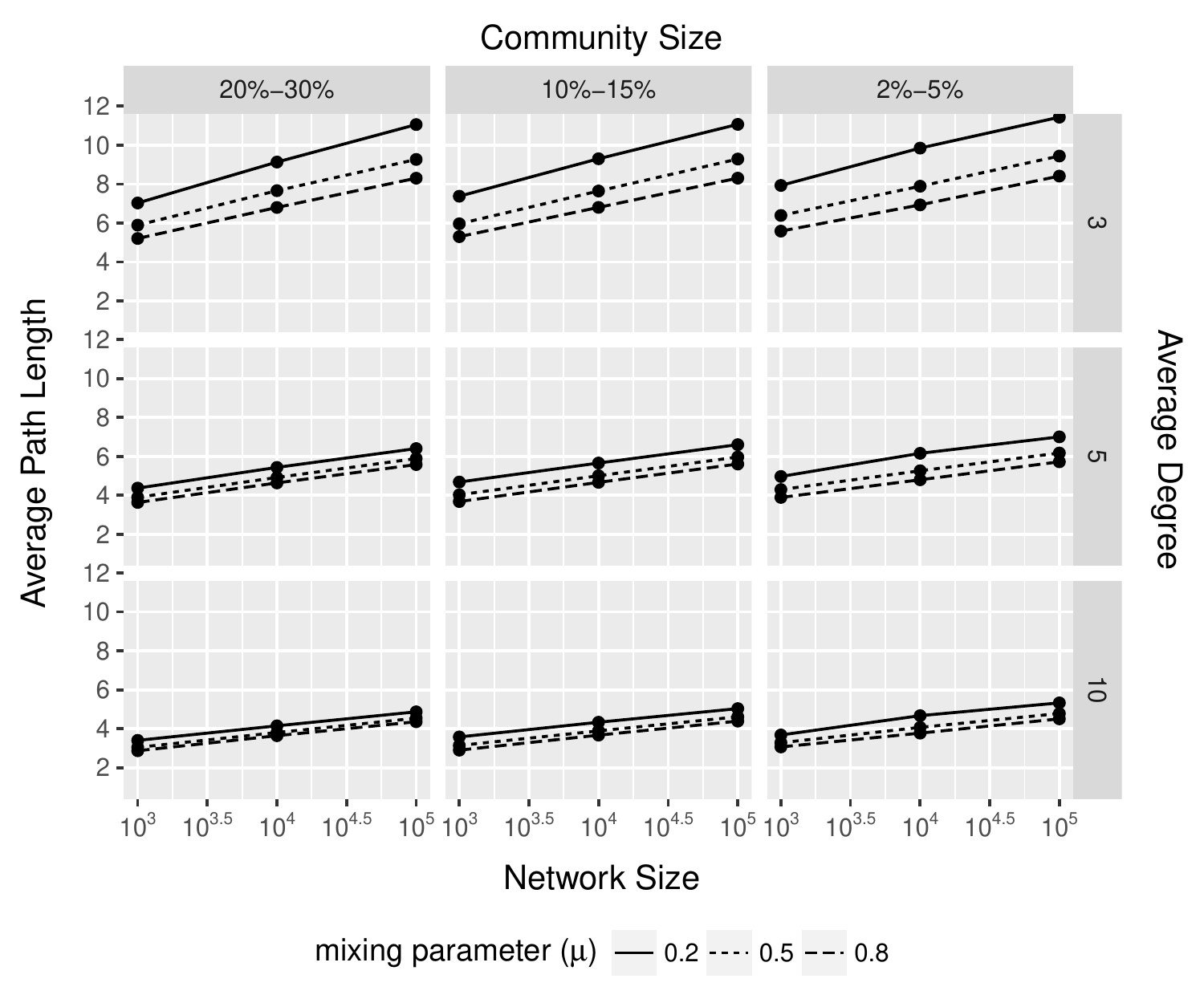}
	\caption{Relationship between average path length and mixing parameter($\mu$) as a function of network size. Results are shown for different community sizes and average node degree. Low average path lengths can be clearly observed for all the networks for various parameters except for slightly higher average path lengths when for low average node degree.}
	\label{fig::newmodel-apl}
\end{figure}

Figure \ref{fig::newmodel-apl} shows the relationship between average path length and mixing parameter ($\mu$) as a function of size of network . The average path length grows logarithmically with respect to the network size regardless of mixing parameter. However, average degree has a direct relation with average path length, as degree increases, the average path length of the network decreases.

The basic premises of the algorithm is generating sub-graphs for given size and then connect these isolated sub-graphs by adding edges among them as per given mixing parameter ($\mu$). 

The algorithm start with two parameters: (1) a list contains $L$ elements where each element represents size of a community and (2) a mixing parameter $\mu$. First, algorithm generates $L$ sub-graphs for given list by using Barabasi-Albert model. At the moment, these graphs are isolated disconnected to each other. Then we calculate require number of inter-cluster edges to be added for each sub-graph. $E_{R}[j]$ represents required number of inter-cluster edges required by $j_{th}$ sub-graph. 

To create inter-cluster edges we adopted similar method as Molly for generating random graphs for given degree sequence \cite{molloy95}. We first select a random node from community with minimum number of required inter-cluster edges (c1) i.e. min($E_{R}$) and  connect it with a node selected randomly from a community which is selected on the basis of preferentially (c2) i.e. community with higher number of required inter-cluster edges will be having more chances to select. On creation of edge, we update the require number of inter-cluster edges for respective sub-graph (cluster). We repeat this process of inter-cluster edges creation until we created all required-inter cluster edges.

\end{document}